\documentclass[aps,prc,reprint,superscriptaddress,nofootinbib,floatfix,showpacs]{revtex4-1}
\pdfoutput=1

\usepackage{epsfig}
\usepackage{dcolumn}
\usepackage{bm}
\usepackage{amssymb}
\usepackage{amsmath}
\usepackage{hyperref}
\usepackage[dvipsnames]{xcolor}
\usepackage{subcaption}

\newcommand{\ord}[1]{\mathcal{O} \left( #1 \right)}
\newcommand{\XX}[4]{ {}^{#1}\mathrm{#2} \, {}^{#3}\mathrm{#4} }

\begin{document}
\title{Ultracentral Collisions of Small and Deformed Systems at RHIC: \\ $\XX{}{U}{}{U}$,	$\XX{}{d}{}{Au}$, $\XX{9}{Be}{}{Au}$, $\XX{9}{Be}{9}{Be}$, $\XX{3}{He}{3}{He}$, and $\XX{3}{He}{}{Au}$ Collisions  }
\author{J. Noronha-Hostler}
\email[Email: ]{jacquelyn.noronhahostler@rutgers.edu}
\affiliation{Department of Physics and Astronomy, Rutgers University, Piscataway, NJ USA 08854}
\author{N. Paladino}
\email[Email: ]{noah.paladino@rutgers.edu}
\affiliation{Department of Physics and Astronomy, Rutgers University, Piscataway, NJ USA 08854}
\author{S. Rao}
\email[Email: ]{svr34@scarletmail.rutgers.edu}
\affiliation{Department of Physics and Astronomy, Rutgers University, Piscataway, NJ USA 08854}
\author{Matthew D. Sievert}
\email[Email: ]{matthew.sievert@rutgers.edu}
\affiliation{Department of Physics and Astronomy, Rutgers University, 
Piscataway, NJ USA 08854}
\author{Douglas E. Wertepny}
\email[Email: ]{wertepny@post.bgu.ac.il}
\affiliation{Department of Physics, Ben-Gurion University of the Negev, Beer-Sheva 84105, Israel}

\date{\today}
\begin{abstract}
	
In this paper, we study a range of collision systems involving deformed ions and compare the elliptic and triangular flow harmonics produced in a hydrodynamics scenario versus a color glass condensate (CGC) scenario.  For the hydrodynamics scenario, we generate initial conditions using TRENTO and work within a linear response approximation to obtain the final flow harmonics.  For the CGC scenario, we use the explicit calculation of two-gluon correlations taken in the high-$p_T$ ``(semi)dilute-(semi)dilute'' regime to express the flow harmonics in terms of the density profile of the collision.  We consider ultracentral collisions of deformed ions as a testbed for these comparisons because the difference between tip-on-tip and side-on-side collisions modifies the multiplicity dependence in both scenarios, even at zero impact parameter.  We find significant qualitative differences in the multiplicity dependence obtained in the initial conditions+hydrodynamics scenario and the CGC scenario, allowing these collisions of deformed ions to be used as a powerful discriminator between models.  We also find that sub-nucleonic fluctuations have a systematic effect on the elliptic and triangular flow harmonics which are most discriminating in $0-1\%$ ultracentral symmetric collisions of small deformed ions and in $0-10\%$ $\XX{}{d}{197}{Au}$ collisions.  The collision systems we consider are $\XX{238}{U}{238}{U}$, $\XX{}{d}{197}{Au}$, $\XX{9}{Be}{197}{Au}$, $\XX{9}{Be}{9}{Be}$, $\XX{3}{He}{3}{He}$, and $\XX{3}{He}{197}{Au}$. 
\end{abstract}

\maketitle

\section{Introduction}

Since the first confirmation of the existence of the Quark Gluon Plasma (QGP) in the early 2000's, many new and unexpected properties have been discovered.  Contrary to expectations, the QGP behaves as a nearly perfect fluid with nearly vanishing shear viscosity to entropy density ratio that is in the range of estimates from AdS/CFT \cite{Niemi:2015voa,Noronha-Hostler:2015uye,Eskola:2017bup,Giacalone:2017dud,Gale:2012rq,Bernhard:2016tnd}.  Heavy-ion collision physicists originally thought that the largest possible ions would be needed to recreate this deconfined state of matter.  However, more recent experiments have surprisingly discovered signals of the fluid-like nature of a QGP droplet in small systems comparable to the size of a proton (seen in pPb, pAu, dAu, $\XX{3}{He}{}{Au}$, and possibly even pp collisions) \cite{Chatrchyan:2013nka,Aaboud:2017acw,Aaboud:2017blb,Aad:2013fja,Sirunyan:2018toe,Chatrchyan:2013nka,Khachatryan:2014jra,Khachatryan:2015waa,Khachatryan:2015oea,Sirunyan:2017uyl,ABELEV:2013wsa,Abelev:2014mda,Adare:2013piz, Adare:2014keg,Aidala:2018mcw,	Adare:2018toe,	Adare:2015ctn,	Aidala:2016vgl,	Adare:2017wlc,	Adare:2017rdq,	Aidala:2017pup,	Aidala:2017ajz}. 

While the predictions of relativistic hydrodynamics have been reasonably consistent with experimental data  \cite{Bozek:2011if,Bozek:2012gr,Bozek:2013ska,Bozek:2013uha,Kozlov:2014fqa,Zhou:2015iba,Zhao:2017rgg,Mantysaari:2017cni,Weller:2017tsr,Zhao:2017rgg}, alternative pictures that do not rely on a tiny QGP appearing in small systems have emerged \cite{Greif:2017bnr,Schenke:2016lrs,Mantysaari:2016ykx,Albacete:2017ajt}. Probably the strongest contender is the color glass condensate (CGC) effective theory, in which the proliferation of small-$x$ gluons can produce many-body correlations and flow-like signals similar to relativistic hydrodynamics.   A number of experiments have been conducted (and future ones proposed \cite{Citron:2018lsq,Sievert:2019zjr,Lim:2018huo,Huang:2019tgz,Giannini:2019abh}) to disentangle these two scenarios; however, as of yet a strong smoking gun signal is still lacking.  

During this time, there has also been significant development in understanding of the influence of nuclear structure on relativistic heavy-ion collisions \cite{Adamczyk:2015obl,wang:2014qxa,Moreland:2014oya,Goldschmidt:2015kpa,Giacalone:2017dud,Schenke:2019ruo}.  Ultracentral collisions can be used as a probe of nuclear deformations, as recently shown for both $\XX{129}{Xe}{129}{Xe}$ \cite{Giacalone:2017dud,CMS:2018jmx,Acharya:2018ihu,ATLAS:2018iom} and $\XX{238}{U}{238}{U}$ collisions  \cite{Adamczyk:2015obl,wang:2014qxa,Moreland:2014oya,Goldschmidt:2015kpa,Schenke:2019ruo}.  In hydrodynamics, a quadrupole deformation of the nucleus enhances the elliptic flow signals in ultracentral collisions where the nuclei are almost entirely overlapping.  Additionally, if one selects on the $0-1\%$ most central collisions using the ZDC at STAR (i.e. events with the fewest number of spectator nucleons), a sensitivity is seen to tip-to-tip vs. side-to-side collision geometries  \cite{Adamczyk:2015obl}.  Tip-to-tip collisions produce the smallest elliptic flow but the largest multiplicity, while in contrast side-to-side collisions produce a larger elliptic flow but a smaller multiplicity.  This anticorrelation between the magnitude of elliptic flow and the multiplicity arises naturally from the eccentricities, which is then seen in the final flow data due to nearly perfect linear response in ultracentral collisions.  We will refer to this as the initial conditions+hydrodynamics scenario.  In this scenario, particles are emitted from the freeze-out hypersurface independently, resulting  in the mutual correlation of bulk particles through the dependence of the single-particle distribution on the event geometry.

In contrast to the hydrodynamic picture where the response is driven by the shape of the collision geometry, gluon correlations in a color glass condensate picture are driven instead by the multiplicity via a dependence on the saturation scales of the colliding ions.  Thus, in a color glass condensate scenario, the collision of two deformed ions would result in entirely the opposite behavior: a positive correlation of elliptic flow with multiplicity.  Tip-on-tip collisions would produce the greatest two-particle correlations owing to the largest saturation scales and multiplicity, while side-on-side collisions would produce the least for the same reasons \cite{Kovchegov:2013ewa}.  The ability to distinguish between these qualitatively opposite dependences on the multiplicity makes ultracentral collisions of deformed ions a powerful tool to study the origin of correlations.  In contrast, for truly round systems, such as ultracentral $\XX{}{p}{}{Au}$ or $\XX{}{Au}{}{Au}$, the initial conditions+hydrodynamics response leads to a flattened dependence on multiplicity in the absence of a geometry bias.

One should be clear that, aside from the above scenario where the final-state correlations are derived entirely from CGC dynamics, it is also possible to base the initial conditions on a CGC picture where the obtained energy-momentum tensor is fed into relativistic hydrodynamics.  The resulting correlations will still fit into the initial condition+hydrodynamics scenario because the hydrodynamic response will still reflect the geometry of the initial energy-momentum tensor via independent particle emission at freezeout.

Generally, the temperatures and densities achieved in central symmetric collisions of large ions such as uranium are expected to produce a QGP, so they are not natural candidates for explanations based on CGC correlations alone.  Thus, in this paper we study smaller ion-on-ion collisions producing smaller multiplicities, such that either scenario could be viable.  Here we explore both symmetric collisions such as $\XX{9}{Be}{9}{Be}$ or $\XX{3}{He}{3}{He}$ and asymmetric collisions such as $\XX{}{d}{197}{Au}$ and $\XX{3}{He}{}{Au}$.  In these small but deformed systems, we see that the prediction of opposite correlations between elliptic flow and multiplicity for two scenarios persists.  We thus argue that deformed ion-ion collisions may be a valuable testbed to discriminate between correlations arising from the color glass condensate vs. initial condition+hydrodynamics scenarios. 

Previous UU hydrodynamic predictions were made in \cite{Moreland:2014oya,Goldschmidt:2015kpa,Schenke:2019ruo} some of which we will discuss and compare with below.  Additionally,  photons \cite{Dasgupta:2016qkq} and  eccentricities \cite{Giacalone:2018apa} in terms of multiparticle cumulants have also been studied.

The rest of this paper is organized as follows.  In Sec.~\ref{sec:Calculation} we set up the calculation, detailing the sampling of initial conditions in Sec.~\ref{ss:IC}, the CGC picture in Sec.~\ref{ss:CGC}, and the initial conditions+hydrodynamics picture in Sec.~\ref{ss:hydro}.  In Sec.~\ref{sec:Results} we present our results for a variety of collision systems, including 
$\XX{238}{U}{238}{U}$ in Sec.~\ref{ss:UU}, $\XX{}{d}{197}{Au}$ in Sec.~\ref{ss:dAu}, $\XX{9}{Be}{197}{Au}$ and $\XX{9}{Be}{9}{Be}$ in Sec.~\ref{ss:Be}, and $\XX{3}{He}{3}{He}$ and $\XX{3}{He}{197}{Au}$ in Sec.~\ref{ss:He}.  Finally we summarize our conclusions in Sec.~\ref{Sec:Conclusions}.  We also detail the particular centrality binning method we use to select the $0-1\%$ most ultracentral collisions in Appendix~\ref{sec:center}, the shape parameters used to describe ${}^{238} \mathrm{U}$ in Appendix~\ref{sec:Def}, and an analysis of the multiplicity dependence in the CGC picture in Appendix \ref{app:CGCmult}.

\section{Calculation of azimuthal anisotropies}
\label{sec:Calculation}

\subsection{Sampling of Multiplicity and Nucleon Positions} 
\label{ss:IC}

All collisions here are generated by an adapted version of TRENTO 2.0 \cite{Moreland:2014oya}.  Within TRENTO the total multiplicity is calculated using
\begin{equation} \label{e:mult1}
S=c \int d^2 x_\bot \: f\Big(T_A (\vec{x}_\bot) , T_B (\vec{x}_\bot)\Big)
\end{equation}
where $T_A , T_B$ are the nuclear profile functions of colliding ions $A$ and $B$ and c is a phenomenological scaling constant that can depend on collision species and center-of-mass energy, which can be fixed by the total particle yields.  For this paper, the precise value of $c$ will be irrelevant since we will only consider quantities where it cancels out.  The effective reduced thickness function $f(A,B)$ can be chosen phenomenologically or determined from theoretical considerations; we consider the two scenarios
\begin{subequations}	\label{e:mult2}
\begin{align}	
f(A,B)&=\sqrt{A B} \qquad (p=0)	\label{e:sqrtTaTb} \\
f(A,B)&=AB \label{e:TaTb},
\end{align}
\end{subequations}
where the $\sqrt{AB}$ configuration is one of the default ``$p=0$'' options in TRENTO, and we have hard-coded a new linear scaling to reflect the predicted scaling of the initial energy density in the color glass condensate \cite{Lappi:2006hq}.  For this paper we vary the size of the nucleon Gaussian width from $\sigma=0.3-0.51$ fm (from \cite{Giacalone:2017uqx} we expect a preference for $\sigma=0.3$ fm in small systems) and we vary the multiplicity scaling factor $k$ from $k=0.5-2$ for systems that have not yet been run yet.  In systems that have already been run we can fix k by direct comparisons to the measured multiplicity distributions.  For details of the TRENTO parameters such as $\sigma$, $k$ and $c$, see \cite{Moreland:2014oya,Bernhard:2015hxa,Bernhard:2016tnd,Giacalone:2017uqx}.

The changes that we have made to TRENTO are as follows:
\begin{itemize}
	\item Inclusion of an option for linear $T_AT_B$ scaling \eqref{e:TaTb}
	\item Addition of the ions $\XX{}{}{9}{Be}$, $\XX{}{}{3}{He}$, and new parameterizations of $^{238}U$.  
	\item Inclusion in the output of the moments $\mathcal{I}_\alpha$ of the nuclear profile functions for $\alpha=1-3$ (See Eq.\ (\ref{e:Integrals})
	\item Addition of the subroutine ``vUSPhydro" to provide compatibility with Lagrangian hydrodynamic codes. 
\end{itemize}
We note that in the case of $\XX{}{}{3}{He}$ \cite{Carlson:1997qn} we have made a .hdf configuration file that can be run within TRENTO as well. 

The v2.0 release of TRENTO contains an option for including sub-nucleonic fluctuations in the density profiles with variable parameters.  In various collision systems considered in this paper, we will study the effect of turning on these sub-nucleonic fluctuations using the out-of-the-box parameters which were previously tuned to $\XX{}{p}{208}{Pb}$ collisions \cite{Moreland:2018gsh}
\footnote{
Note that some values quoted here have changed slightly from Ref.~\cite{Moreland:2018gsh}, as per private communications from the authors.
}
.  Those parameters are as follows: $p=0, k=0.19, n_c = 6, w=0.855~\mathrm{fm}, r_{cp} = 0.81~\mathrm{fm}, v=0.43~\mathrm{fm}, d_{min} = 0.81~\mathrm{fm}$.

\subsection{Two-Gluon Correlations from the CGC}
\label{ss:CGC}

Mirroring the notation of Ref.~\cite{Luzum:2013yya}, we can  decompose the two-particle multiplicity in a given event as
\begin{align} \label{e:2particle}
& \frac{dN_2}{d^2 p_1 dy_1 \, d^2 p_2 dy_2} =
\frac{dN_1}{d^2 p_1 dy_1} \, \frac{dN_1}{d^2 p_2 dy_2} +
\delta_2 (p_1 , p_2)
\end{align}
Here the first term, identified as ``collective flow,'' arises from the consequences of coupling the single-particle distribution $\frac{dN_1}{d^2p \, dy}$ to characteristic directions in the collision, such as the orientation of the collision geometry or the direction of the color electric field in a color domain \cite{Dumitru:2014yza, Dumitru:2015cfa}.  The second term $\delta_2$ describes the ``non-flow'' terms arising from genuine correlations between the particles, such QCD scattering. The two-particle cumulant is defined as
\begin{align}	\label{e:vn2def_1}
\left( v_n \{2\} \right)^2 &\equiv 	
\left\langle e^{i n (\phi_1 - \phi_2)} \right\rangle
\notag \\ &=
\frac{
\left\langle
\int_{p_1 p_2} e^{i n (\phi_1 - \phi_2)} \,
\frac{dN_2}{d^2 p_1 \, dy_1 \: d^2 p_2 \, dy_2}
\right\rangle
}
{\langle N_{\mathrm{pairs}} \rangle} .
\end{align}
It is important to note that, while the average defined in Eq.~\eqref{e:vn2def_1} is often written as an average of the ratio:
\begin{align*}
	\left\langle
	\frac{1}{N_{\mathrm{pairs}}}
	\int_{p_1 p_2} e^{i n (\phi_1 - \phi_2)} \,
	\frac{dN_2}{d^2 p_1 \, dy_1 \: d^2 p_2 \, dy_2}
	\right\rangle ,
\end{align*}
in practice it is actually computed as in Eq.~\eqref{e:vn2def_1}.  The difference between these quantities is embodied in the correlated fluctuations of multiplicity in tandem with the azimuthal anisotropies.  For small enough multiplicity bins, these differences in the two-particle sector may be numerically small, but they are conceptually different quantities and can affect the systematics of higher-order cumulants. 

In particular, note that the definition of what is considered ``an event'' in \eqref{e:2particle} is arbitrary and potentially model-dependent.  In some CGC calculations, for instance, a random sampling of color sources $\rho$ is performed such that the ``event by event fluctuations'' include the fluctuations of these color fields.  In others, the sampling is performed only at the level of the fluctuating event geometry, with the color fields already having been averaged to obtain color multipole operators, such as Wilson line dipoles and quadrupoles.  These various (model-dependent) distinctions can matter when computing the average of the ratio, but for the experimentally relevant quantity \eqref{e:vn2def_1}, the definition of what constitutes ``an event'' is arbitrary and does not affect the result for $v_n \{2\}$.

Expanding \eqref{e:vn2def_1} using the parameterization \eqref{e:2particle} gives
\begin{align}	\label{e:vn2def_2}
\left( v_n \{2\} \right)^2 &=
\frac{
	\left\langle \left|
	\int_{p} e^{i n \phi} 
	\frac{dN_1}{d^2 p \, dy}
	\right|^2 \right\rangle
}
{\langle N_{\mathrm{pairs}} \rangle}
\notag \\ & \hspace{0.5cm} +
\frac{
	\left\langle
	\int_{p_1 p_2} e^{i n (\phi_1 - \phi_2)} \,
	\delta_2 (p_1 , p_2)
	\right\rangle
}
{\langle N_{\mathrm{pairs}} \rangle} ,
\end{align}
where the first term arises from ``flow'' and the second arises from ``non-flow.''  For the present purposes, we consider an ``event'' in the case of two-gluon production in the CGC to be one randomly sampled collision geometry.  Then in one event, the single-gluon distribution is isotropic
\begin{align}
\frac{dN_1}{d^2 p \, dy} = \frac{1}{2\pi p_T} \frac{dN_1}{dp_T \, dy}
\end{align}
such that the first (``flow'') term of \eqref{e:vn2def_2} vanishes.  Then $v_n\{2\}$ is obtained entirely due to the genuine two-particle correlations $\delta_2$.

As derived in Ref.~\cite{Kovchegov:2012nd}, when the transverse momenta are much larger than the saturation scale $p_T^2 \gg Q_s^2$, the CGC correlations between two gluons emitted in heavy-light ion collisions can be written as:
\begin{align} \label{e:deltatwo2}
\delta_2 (p_1, p_2) \overset{\mathrm{L.O.}}{=} \left( \int d^2 x_\bot T_A^2 (\vec{x}_\bot) \, T_B^2 (\vec{x}_\bot) \right) \: f (p_1 , p_2) .
\end{align}	
This statement is true at lowest order in the (semi)dilute-dense power counting, and $f$ is a known function of the transverse momenta of the gluons which is manifestly symmetric under $\vec{p}_{2 \bot} \rightarrow - \vec{p}_{2 \bot}$, corresponding to $\Delta \phi \rightarrow \Delta \phi + \pi$.  As such, \eqref{e:deltatwo2} does not contribute to $v_n \{2\}$ for odd values of $n$.

Similarly, at the next order in the (semi)dilute-dense CGC power counting, the two-gluon correlations expressed in Eq. (77) of Ref.~\cite{Kovchegov:2018jun} have a comparable form,
\begin{align} \label{e:deltatwo3}
\delta_2 (p_1, p_1) \overset{\mathrm{N.L.O.}}{=} \left( \int d^2 x_\bot T_A^3 (\vec{x}_\bot) \, T_B^3 (\vec{x}_\bot) \right) \: g (p_1 , p_2),
\end{align}	
but at this order the function $g(p_1, p_2)$ breaks the $\Delta \phi \rightarrow \Delta \phi + \pi$ symmetry present at lowest order.  Thus Eq.~\eqref{e:deltatwo3} provides the first nonvanishing contribution to the cumulants $c_n \{2\}$ and $c_n \{4\}$ for odd values of $n$.  
\footnote{We note that the expressions \eqref{e:deltatwo2} and \eqref{e:deltatwo3} are derived under the assumption that transverse gradients of the nuclear profile functions are small compared to perturbative length scales $\ll 1/\Lambda_{\mathrm{QCD}}$.  This assumption is clearly valid for a smooth nuclear profile function, but for a Monte Carlo sampling with nucleon-scale fluctuations, its validity is far less clear.}

Thus, in a given event, the two-particle azimuthal modulation for even harmonics $2n$ is given by
\begin{align}
\left( v_{(2n)} \{2\} \right)^2 &= 
\frac{1}{\langle N_{\mathrm{pairs}} \rangle}
\left\langle \int d^2 x_\bot T_A^2 (\vec{x}_\bot) \, T_B^2 (\vec{x}_\bot) \right\rangle
\notag \\ & \hspace{-0.25cm} \times
\left[ \int_{p_1 p_2} \:
e^{i (2n) (\phi_1 - \phi_2)} \:
f (p_1 , p_2)
\right]
\notag \\ \notag \\ & \hspace{-0.25cm} \equiv
\frac{1}{\langle N_{\mathrm{pairs}} \rangle}
\left\langle \int d^2 x_\bot T_A^2 (\vec{x}_\bot) \, T_B^2 (\vec{x}_\bot) \right\rangle
\times f_{(2n)} ,
\end{align}
where all the event-by-event fluctuations are contained in the geometry profiles $T_{A, B} (\vec{x}_\bot)$ and the azimuthal modulation $f_{(2n)}$ is independent of the event.  Note that all dependence on which even harmonic $(2n)$ is chosen is also contained entirely within the constant $f_{(2n)}$.  
Likewise, for the odd harmonics $(2n-1)$, the event-by-event azimuthal modulation takes the form
\begin{align}
\left( v_{(2n-1)} \{2\} \right)^2 &= 
\frac{1}{\langle N_{\mathrm{pairs}} \rangle}
\left\langle \int d^2 x_\bot T_A^3 (\vec{x}_\bot) \, T_B^3 (\vec{x}_\bot) \right\rangle
\notag \\ & \hspace{-1cm} \times
\left[ \int_{p_1 p_2} \:
e^{i (2n-1) (\phi_1 - \phi_2)} \:
g (p_1 , p_2)
\right]
\notag \\ \notag \\ & \hspace{-1cm} \equiv
\frac{1}{\langle N_{\mathrm{pairs}} \rangle}
\left\langle \int d^2 x_\bot T_A^3 (\vec{x}_\bot) \, T_B^3 (\vec{x}_\bot) \right\rangle
\times g_{(2n-1)} .
\end{align}
While a complete determination of $f_{(2n)}$ and $g_{(2n-1)}$ are necessary for the absolute normalization or the $p_T$ dependence of the cumulants $v_n \{2\}$, they cancel out completely for the $p_T$-integrated quantities considered here when expressed in terms of ratios.  We can express the even and odd harmonics in the same notation by defining
\begin{align}
\nu_n \equiv (n \, \mathrm{Mod} \, 2) + 2=
\begin{cases}
	2 \hspace{0.5cm} \mathrm{if} \: $n = $\mathrm{even}
	\\
	3 \hspace{0.5cm} \mathrm{if} \: $n = $\mathrm{odd}
\end{cases} .
\end{align} 
Then the relevant quantities are the integrals of the nuclear densities $T_{A/B}$ in a given event, raised to some power $\alpha$:
\begin{align} \label{e:Integrals}
\mathcal{I}_\alpha \equiv \int d^2 x_\bot 
T_A^\alpha (\vec{x}_\bot) T_B^\alpha (\vec{x}_\bot) .
\end{align}

Now we would like to use these quantities to compare the cumulants $v_n \{2\}$ for ultracentral collisions, e.g. $0-1\%$ centrality determined from the ZDC, versus a smaller re-binning of those same events e.g. in $20$ sub-bins sorted by multiplicity.  We will express the ratio of the cumulants in an individual sub-bin $i$ to the full ultracentral selection cut as $v_n^i \{2\} / v_n \{2\}$, which we can express as
\begin{align} \label{e:cum2}
\frac{	v_n^i \{2\}	}{	v_n \{2\}	}	&= 
\sqrt{
\frac{	\Big\langle N_{\mathrm{tot}}^2 	\Big\rangle_{0-1\%}	}
	{	\Big\langle N_{\mathrm{tot}}^2 	\Big\rangle_{i}	}
\:
\frac{	\Big\langle \mathcal{I}_{\nu_n}	\Big\rangle_{i}	}
	{	\Big\langle \mathcal{I}_{\nu_n}	\Big\rangle_{0-1\%}	} 
},
\end{align}
where we have approximated $N_{\mathrm{pairs}} = N_{\mathrm{tot}} (N_{\mathrm{tot}} - 1) \approx N_{\mathrm{tot}}^2$.  The multiplicity $N_{\mathrm{tot}}$ is determined in TRENTO from Eq.~\eqref{e:mult1} from one of the parameterizations \eqref{e:mult2} of the reduced thickness function.  For both cases the multiplicity is proportional to one of the weighted integrals \eqref{e:Integrals} as 
\begin{align}
N_{\mathrm{tot}} \propto \mathcal{I}_r ,
\end{align}
with the exponent $r = 1/2$ corresponding to the the phenomenological ``$p=0$'' setting \eqref{e:sqrtTaTb} and $r = 1$ corresponding to the new hard-coded configuration \eqref{e:TaTb} according to the scaling of the initial energy density in the CGC.  With these expressions for the multiplicity, the cumulants \eqref{e:cum2} are now entirely controlled by moments of the nuclear profiles \eqref{e:Integrals}:
\begin{align}	\label{e:cum3v2}
\frac{	v_n^i \{2\}	}{	v_n \{2\}	}	&= 
\sqrt{
	\frac{	\Big\langle (\mathcal{I}_r)^2 	\Big\rangle_{0-1\%}	}
	{	\Big\langle (\mathcal{I}_r)^2 	\Big\rangle_{i}	}
	\:
	\frac{	\Big\langle \mathcal{I}_{\nu_n}	\Big\rangle_{i}	}
	{	\Big\langle \mathcal{I}_{\nu_n}	\Big\rangle_{0-1\%}	} 
} .
\end{align}
We then generate a large number of collision geometries for a given choice of ion species, select the ultracentral $0-1\%$ bin and a set of sub-bins $i$, and use \eqref{e:cum3v2} to determine the multiplicity dependence of the cumulants in the CGC framework.

\subsection{Collective Flow in Hydrodynamics} 
\label{ss:hydro}

Hydrodynamics is, at its core, an initial value problem, with a key input (and a significant source of uncertainty) being the initial conditions for the hydrodynamic evolution.  Most models employ some sort of wounded nucleon picture, including the event-by-event fluctuations of nucleon positions.  In its most simplistic form, this corresponds to a Glauber model where the nucleon positions are sampled from a Wood-Saxon density distribution 
\begin{equation}\label{eqn:wood}
\rho=\rho_0\left[1+\exp\left(\frac{r-R(\theta)}{a}\right)\right]^{-1} 
\end{equation}
where in the case of a deformed nucleus, the radius is not a constant but varies with azimuthal angle:
\begin{equation}\label{eqn:rad_def}
R=R_0\left(1+\underbrace{\beta_2  Y_{20}(\theta)}_{Quadrupole}+\underbrace{\beta_4 Y_{40}(\theta)}_{Hexadecapole} + \cdots \right)
\end{equation}
More sophisticated models include some kind of pre-equilibrium dynamics, such as the classical Yang-Mills evolution of color fields employed in the IP-Glasma \cite{Gale:2012rq} code, and still others like
TRENTO \cite{Moreland:2014oya,Bernhard:2016tnd} take an agnostic approach instead use phenomenological parameterizations combining the nuclear thickness functions.  For the proton, deuteron, and $\XX{}{}{3}{He}$ a Woods-Saxon distribution is not appropriate; we instead sample the nucleon positions using the Hulth\`en wave function \cite{Hulten:1957} for the deuteron and the nucleonic configurations from Ref.~\cite{Carlson:1997qn} for $\XX{}{}{3}{He}$.

The above prescriptions describe the procedure for sampling the positions of nucleons in the colliding nuclei, which are traditionally the degrees of freedom used to determine the deposition of energy or entropy in the initial conditions of hydrodynamics.  More microscopic treatments of the initial conditions, however, can also include fluctuations of nucleonic substructure which mimic in some fashion the partonic content of the nucleon.  In an updated version of TRENTO \cite{Moreland:2018gsh}, for instance, nucleonic substructure is implemented and calibrated to $\XX{}{p}{}{Pb}$ and $\XX{}{Pb}{}{Pb}$ collisions at LHC energies.  As a way to test the sensitivity to these subnucleonic fluctuations, we compare the results obtained by using standard nucleon degrees of freedom with an out-of-the-box application of TRENTO's subnucleonic fluctuations to RHIC at $\sqrt{s_{NN}}=200$ GeV.  

We emphasize one caveat to this comparison: one may expect differences both in the multiplicity fluctuations and in the number of constituents at the lower beam energies.  A detailed recalibration to RHIC kinematics would require the dedication of significant computational resources which is beyond the scope of this paper.  Moreover, since we are proposing collisions of new ion species, in some cases there is no data currently available against which such a recalibration could be performed.  Finally, we note that previous papers have explored the question of sub-nucleonic fluctuations from a different angle by systematically smoothing out smaller scale structures, finding almost no effect in the final flow observables \cite{Noronha-Hostler:2015coa,Gardim:2017ruc}.

The eccentricities $\varepsilon_n$ are defined with respect to the center of mass of the initial entropy deposition such that
\begin{equation} \label{e:ecc_n}
\mathcal{E}_n \equiv - \frac{ \langle \int r^n e^{in\phi}s(r,\phi) r dr d\phi \rangle}{\langle \int r^n s(r,\phi) r dr d\phi \rangle} \equiv \varepsilon_n e^{i n \Phi_n}
\end{equation}
with $\mathcal{E}_1 \equiv 0$ by definition and the minus sign by convention.  In order to study the deformation of Uranium we compare a number of different choices of the parameters $a, R_0, \beta_2, \beta_4$ in Eqs.\ (\ref{eqn:wood}-\ref{eqn:rad_def}).  Our choice of parameters are often based upon previous work, which is cited accordingly in Table\ \ref{tab:wood}. To the best of our knowledge, the parameterization used in Refs.~\cite{Moller:2015fba,Giacalone:2018apa} is the most up-to-date according to the low-energy nuclear structure community. 

\begin{table}
\begin{tabular}{lllll}
\hline
a  [fm]& $R_0$ [fm] & $\beta_2$ & $\beta_4$ & Origin \\
\hline
0.5 &  6.61& 0.236 & 0.098 & (this work) \\
0.55 & 6.86 & 0.28 & 0.093 & \cite{Schenke:2019ruo} \\
0.6 &  6.81& 0.236 & 0.098 & \cite{Moller:2015fba,Giacalone:2018apa} \\
0.6 &  6.81& 0.28 & 0.093 &  \cite{Moreland:2014oya} \\
 \hline
\end{tabular}
\caption{Deformed Wood-Saxon parameters for Uranium initial conditions tested in this paper.  TRENTO \cite{Moreland:2014oya} comes with three default settings U (listed above) and U2 and U3, which are not shown.}\label{tab:wood}
\end{table}

\begin{figure}[h]
\centering
\includegraphics[width=1\linewidth]{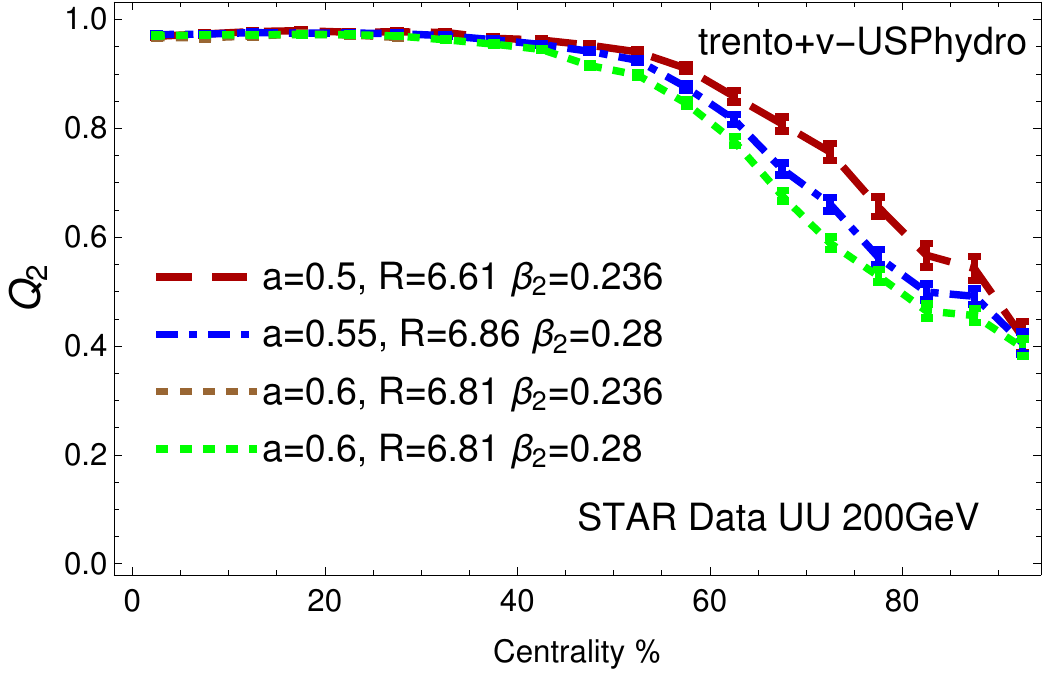}
\caption{(Color online) Pearson coefficient for UU 193 GeV of various deformed Wood Saxon parameterization of Uranium.}
\label{fig:Q2}
\end{figure}

To a good approximation, the single-particle anisotropic flow harmonics in hydrodynamics
\begin{align}
V_n = 
\frac{
	\left\langle \int d^2 p \, dy \, e^{i n \phi} 
	\frac{dN_1}{d^2 p \, dy}
	\right\rangle
}
{\langle N_{\mathrm{tot}} \rangle}
\equiv v_n e^{i n \psi_n}
\end{align}
arise primarily from linear response \cite{Hirano:2010je,Teaney:2010vd,Qiu:2011hf,Qiu:2011iv,Gardim:2011xv,Niemi:2012aj,Teaney:2012ke,Gardim:2014tya,Betz:2016ayq} from the initial eccentricities \eqref{e:ecc_n}:
\begin{align} \label{e:linear}
V_n \approx \kappa_n \mathcal{E}_n 
\end{align}
for some complex coefficient $\kappa_n$.  Non-linear deviations from the linear proportionality \eqref{e:linear} have also been shown to be relevant for the prediction of $v_2\{4\}/v_2\{2\}$ \cite{Noronha-Hostler:2015dbi,Sievert:2019zjr}.   The quality of the approximation \eqref{e:linear} is described by the Pearson coefficient 
\begin{align}
Q_n \equiv\frac{\langle V_n \mathcal{E}_n^* \rangle}
{	
	\sqrt{ \langle | V_n |^2 \rangle} \, 
	\sqrt{\langle | \mathcal{E}_n |^2 \rangle}
} 
\end{align}
with $Q_2 \rightarrow 1$ when linear response is exact.  As shown in Fig.\ \ref{fig:Q2}, linear response \eqref{e:linear} is a nearly perfect approximation to the full hydrodynamic simulation in most central $\XX{238}{U}{238}{U}$ collisions, independent of the geometry parameters.  This comparison was performed by generating multiple different sets of initial conditions within TRENTO and then running them through the event-by-event relativistic viscous hydrodynamic model v-USPhydro \cite{Noronha-Hostler:2013gga,Noronha-Hostler:2014dqa}, using the same hydrodynamic parameters as in \cite{Alba:2017hhe} for $\XX{}{Au}{}{Au}$ collisions.  

In \cite{Sievert:2019zjr} the Pearson coefficient $Q_2$ was also studied versus system size.  It is true that the linear mapping does not hold quite as well for smaller $N_{\mathrm{part}}$.  However, it was still found that smaller systems have a better linear mapping than larger systems with the same $N_{\mathrm{part}}$; that is, a central small system has a better linear mapping than a peripheral larger system.  Since we consider only ultracentral collisions in this paper, it then seems reasonable to assume that linear mapping still holds in ultracentral small systems.  We note that, while deviations from linear response can affect the magnitude of $v_n \{2\}$, they should not strongly affect its multiplicity dependence, and much of the systematic uncertainties associated with assuming linear response will cancel in the ratio $v_n^i \{2\} / v_n \{2\}$.  The accuracy of a linear response assumption can in principle be tested against our full hydrodynamic simulations, but -- as with a recalibration of subnucleonic fluctuations at RHIC -- will require the dedication of significant computational resources. We leave such an analysis for future work, when we hope that some of the new collision species considered here will be approved.

Accordingly, for most of this paper we will assume linear response \eqref{e:linear} in order to calculate high statistics results in ultracentral UU collisions at RHIC.  Running a full hydrodynamics simulation on an event-by-event basis is not realistic since millions of events are required to perform such an analysis.  Then, since we calculate only ratios such as $v_n^i \{2\} / v_n \{2\}$, the coefficient $\kappa_n$ cancels out and we can directly use the eccentricities of the initial conditions to make flow predictions in central collisions. 

In the flow-only scenario, the final-state two-particle distribution \eqref{e:2particle} $\frac{dN_1}{d^2 p \, dy}$ with $\delta_2 = 0$.  Then the second term of \eqref{e:vn2def_2} vanishes, and the two-particle cumulant $v_n \{2\}$ is simply the root-mean-square of the corresponding single-particle flow harmonic:
\begin{eqnarray}\label{eqn:flow_cumulants}
v_n\{2\}&=&\sqrt{\langle v_n^2\rangle} \approx | \kappa_n |^2
\sqrt{\langle \varepsilon_n^2 \rangle}.
\end{eqnarray}
As emphasized previously, this arises because in hydrodynamics the particle are emitted independently at freeze-out.  Correspondingly, the ratio of cumulants in an individual sub-bin $i$ to the full $0-1\%$ ultracentral selection cut is
\begin{align}	\label{e:cum4v2}
\frac{v_n^i \{2\}}{v_n \{2\}} = \sqrt{
\frac{\langle \varepsilon_n^2 \rangle_i}
{\langle \varepsilon_n^2 \rangle_{0-1\%}}
}
\end{align}
which can be compared with the CGC scaling in \eqref{e:cum3v2}.

\section{Results}
\label{sec:Results}

In this Section, we will show results for the ratio of cumulants in sub-bins of ultracentral collisions of various ion species, comparing the predictions of two-gluon correlations from the CGC-only scenario \eqref{e:cum3v2} versus the initial conditions+hydrodynamics scenario (with linear response) \eqref{e:cum4v2}.  In Sec.~\ref{ss:UU} we examine the differences for $\XX{238}{U}{238}{U}$ collisions: a symmetric collision of highly-deformed ions for which the CGC scaling was first demonstrated \cite{Kovchegov:2013ewa}.  In Sec.~\ref{ss:dAu} we study $\XX{}{d}{197}{Au}$ collisions as an example of an asymmetric collision between a small, deformed ion (deuteron) with a large, round ion (gold).  In Sec.~\ref{ss:Be} we study collisions involving an intermediate but highly-deformed ion: ${}^9 \mathrm{Be}$.  We consider both symmetric $\XX{9}{Be}{9}{Be}$ and asymmetric $\XX{9}{Be}{197}{Au}$ collisions.  And in Sec.~\ref{ss:He} we similarly consider the generation of triangular flow $v_3\{2\}$ from collisions involving an inherently triangular nucleus: ${}^3 \mathrm{He}$.  We consider both symmetric $\XX{3}{He}{3}{He}$ and asymmetric $\XX{3}{He}{197}{Au}$ collisions.  Relevant details about the centrality class selection and Wood-Saxon parameters for $\XX{238}{U}{238}{U}$ are contained in Appendices \ref{sec:center} and \ref{sec:Def}.

\subsection{$\XX{238}{U}{238}{U}$ Collisions}
\label{ss:UU}

In Ref.~\cite{Kovchegov:2013ewa}, $\XX{238}{U}{238}{U}$ collisions were considered, leading to the prediction of opposite correlations of $v_n \{2\}$ with multiplicity $M = \frac{dN}{dy}$ between the CGC-mediated two-gluon correlations and the initial conditions+hydrodynamics picture.  While it is not widely believed that CGC mechanisms alone can describe the hadronic correlations in such a large system, it is still interesting to examine a highly-deformed ion like uranium to study the sensitivity of both pictures to deformed nuclei.  This is especially true, given that the relevant data has already been published by the STAR Collaboration \cite{Adamczyk:2015obl} and that $\XX{238}{U}{238}{U}$ is the only collisional system of deformed ions with published data to date for this specific measurement.

\begin{figure}
	\includegraphics[width=\linewidth]{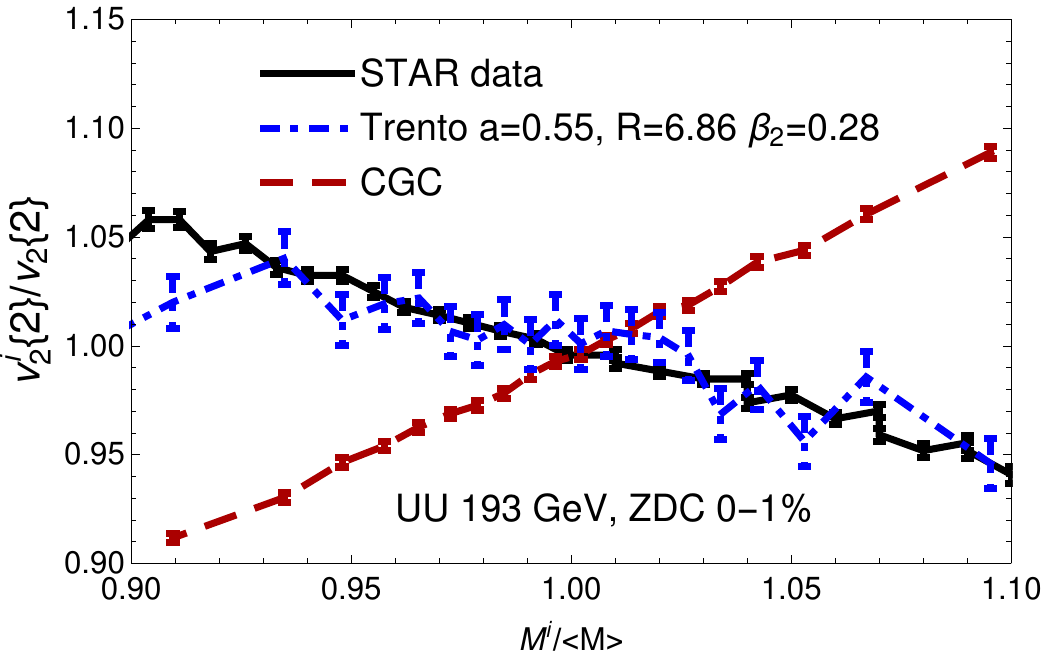}
	\caption{Elliptic flow versus multiplicity for the $0-1\%$ most central $\XX{238}{U}{238}{U}$ collisions by ZDC at RHIC (see Appendix~\ref{sec:center}).  Note that the axes are scaled by the average multiplicity $\langle M\rangle$ and elliptic flow $v_n\{2\}$ for the entire $0-1\%$ centrality class.  Here the multiplicity is determined using the $\sqrt{T_A T_B}$ scaling (the $p=0$ configuration of TRENTO \eqref{e:sqrtTaTb}).}  
	\label{f:ZDCUU}
\end{figure}

The results for $\XX{238}{U}{238}{U}$ are shown in Fig.\ \ref{f:ZDCUU}, where we have subdivided the $0-1\%$ centrality class into 20 sub-bins $i$ by multiplicity.  In contrast to the initial condition+hydrodynamic picture, which leads to an anticorrelation of $v_2^i \{2\}$ with $M^i$, the CGC picture positively correlates the two.  This behavior is qualitatively inconsistent with the STAR data \cite{Adamczyk:2015obl}, indicating (unsurprisingly) that an initial-state only picture cannot describe ultracentral $\XX{238}{U}{238}{U}$ collisions; in contrast, the initial conditions+hydrodynamics picture is able to describe the data with the indicated model parameters (Appendix~\ref{sec:Def}).  

\begin{figure}
	\includegraphics[width=\linewidth]{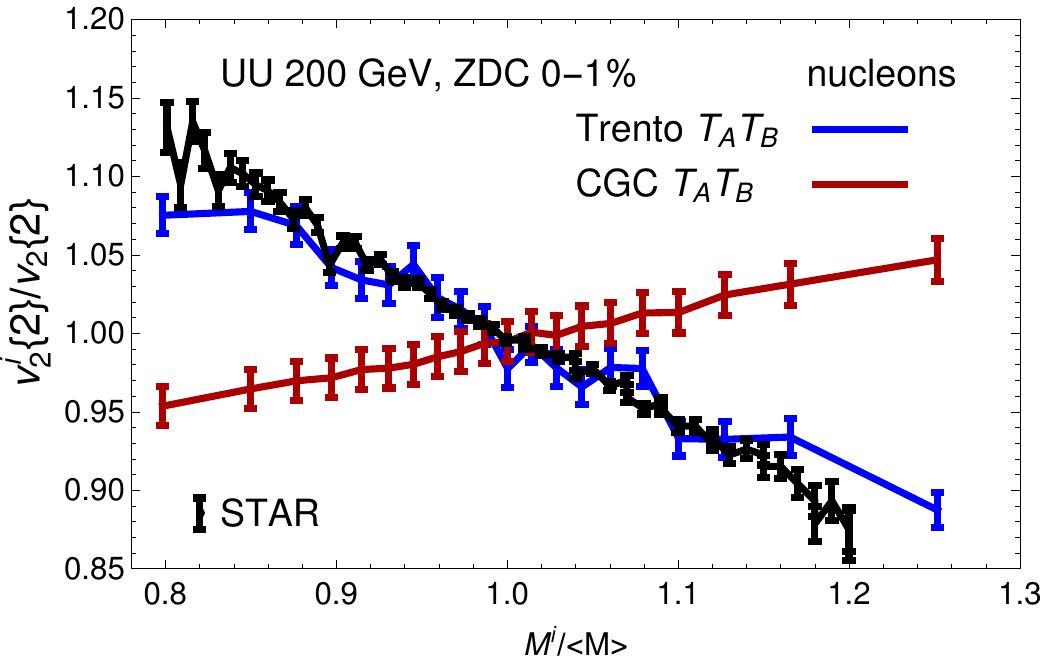}
	\caption{Same as Fig.~\ref{f:ZDCUU} but with linear $T_A T_B$ scaling \eqref{e:TaTb} for the multiplicity.}  
	\label{f:ZDCUU_TaTb}
\end{figure}

%%%%%%This is not correct here with TaTb scaling figs
%	Compare slopes of CGC curve in Fig. 2 vs Fig. 3:
%
%	Fig 2:	(1.09 - 0.91) / (1.09 - 0.91) = 1!
%	Fig 3:	(1.05 - 0.95) / (1.25 - 0.8) = 0.1 / 0.45 = 0.22
%
%%%%%%%%%%%%%%%%%%%%%%%%%%%%

It is important to note that the manner in which the multiplicity dependence is calculated is subject to significant theoretical uncertainty, due in part to the nonperturbative effects associated with hadronization.  To examine this theoretical uncertainty, we compare Fig.~\ref{f:ZDCUU}, which uses a multiplicity determined by the phenomenological $p=0$ configuration of TRENTO \eqref{e:sqrtTaTb} such that $M \propto \int \sqrt{T_A T_B}$, with Fig.~\ref{f:ZDCUU_TaTb}, which uses the linear scaling $M \propto \int T_A T_B$ from the CGC \eqref{e:TaTb}.  One notable effect is the enhancement of multiplicity fluctuations with a $T_A T_B$ scaling, as reflected in the greater range of the horizontal axis.  The initial conditions+hydrodynamics picture is largely unaffected here by the change in multiplicity scaling and is consistent with the STAR data in both Figs.~\ref{f:ZDCUU} and \ref{f:ZDCUU_TaTb}.  The CGC picture, on the other hand, sees the magnitude of its positive slope reduced by switching to $T_A T_B$ multiplicity scaling.

While the disfavoring of an initial-state-only model for a heavy-ion collision like $\XX{238}{U}{238}{U}$ was largely a foregone conclusion, the ability to discriminate between qualitatively different models demonstrated in Fig.~\ref{f:ZDCUU} serves as a baseline proof of concept which we will next extend to smaller systems where the expected outcome is not so clear.

\subsection{$\XX{}{d}{197}{Au}$ Collisions}
\label{ss:dAu}

\begin{figure}[!h]
	\includegraphics[width=\linewidth]{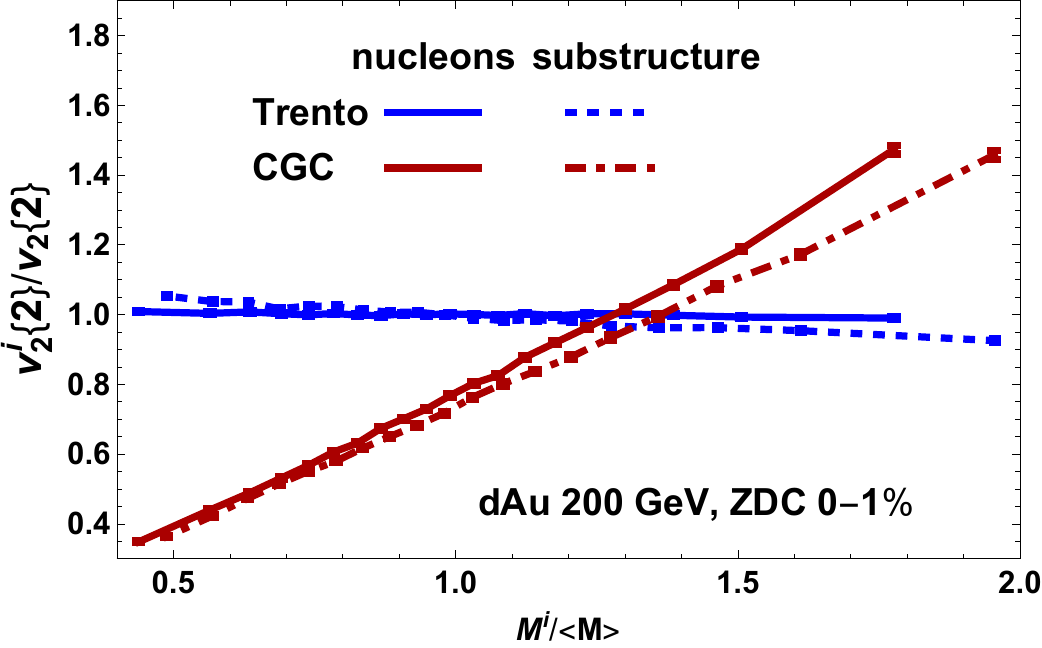}
	\caption{Elliptic flow versus multiplicity for the $0-1\%$ most central $\XX{}{d}{}{Au}$ collisions by ZDC at $200~\mathrm{GeV}$ at RHIC, using the $p=0$ multiplicity scaling \eqref{e:sqrtTaTb} proportional to $\sqrt{T_A T_B}$.  Solid curves indicate nucleon degrees of freedom; dashed curves include $n=6$ sub-nucleonic constituents.} 
	\label{f:ZDCdAu}
\end{figure}

Next we move to a smaller, asymmetric system: deuteron-gold ($\XX{}{d}{}{Au}$) collisions at $\sqrt{s_{NN}}=200$ GeV.  The deuteron is a deformed system, possessing a dominant elliptical geometry; however the proton-neutron separation can be rather wide, colliding as two well-separated nucleons or even with one missing the gold ion entirely.  Very central $\XX{}{d}{}{Au}$ collisions can also correspond to a ``tip-on'' deuteron orientation, with the two nucleons aligned close to the beam axis such that the transverse profile is very round.  The result, shown in the solid curves in Fig.~\ref{f:ZDCdAu}, is that the geometry-driven response of the initial conditions+hydrodynamics picture is nearly flat across multiplicity, whereas the CGC picture still yields elliptic flow which increases with the multiplicity.

The emerging qualitative picture is that hydrodynamic picture and the CGC picture couple oppositely to deformed geometries; however, the quantitative magnitude of this effect can depend on nonperturbative details of the models.  These include not only the choice of multiplicity schemes like $M \propto \int \sqrt{T_A T_B}$ versus $M \propto \int T_A T_B$ as compared in Figs.~\ref{f:ZDCUU} and \ref{f:ZDCUU_TaTb}, but also the choice of whether to consider possible nucleon substructure in the event-by-event fluctuations of the nuclear profile functions.  To study this latter effect, we use the constituent TRENTO model with $n=6$ constituents per nucleon with parameters that were previously tuned to $\XX{}{p}{}{Pb}$ collisions \cite{Moreland:2018gsh}.  The substructure effect is reflected in the dashed curves in Fig.~\ref{f:ZDCdAu}.  We see a modest $\sim\ord{5\%}$ enhancement in $v_2 \{2\}$ in the initial conditions+hydrodynamics picture, reflecting a small enhancement in the overall ellipticity of the deuteron when substructure is included.  In contrast, turning on substructure leads to a slight flattening of the multplicity dependence in the CGC picture.  As discussed in Appendix~\ref{app:CGCmult}, while the gluon correlations leading to $v_2 \{2\}$ in the CGC picture are not directly affected by the ellipticity of the collision geometry, they are affected by the smoothness or lumpiness of the distribution.  Going from a deuteron characterized by nucleon degrees of freedom to one including sub-nucleonic fluctuations results in a somewhat lumpier density distribution, which generally flattens the multiplicity dependence of the CGC effect.

\begin{figure}[!h]
	\includegraphics[width=\linewidth]{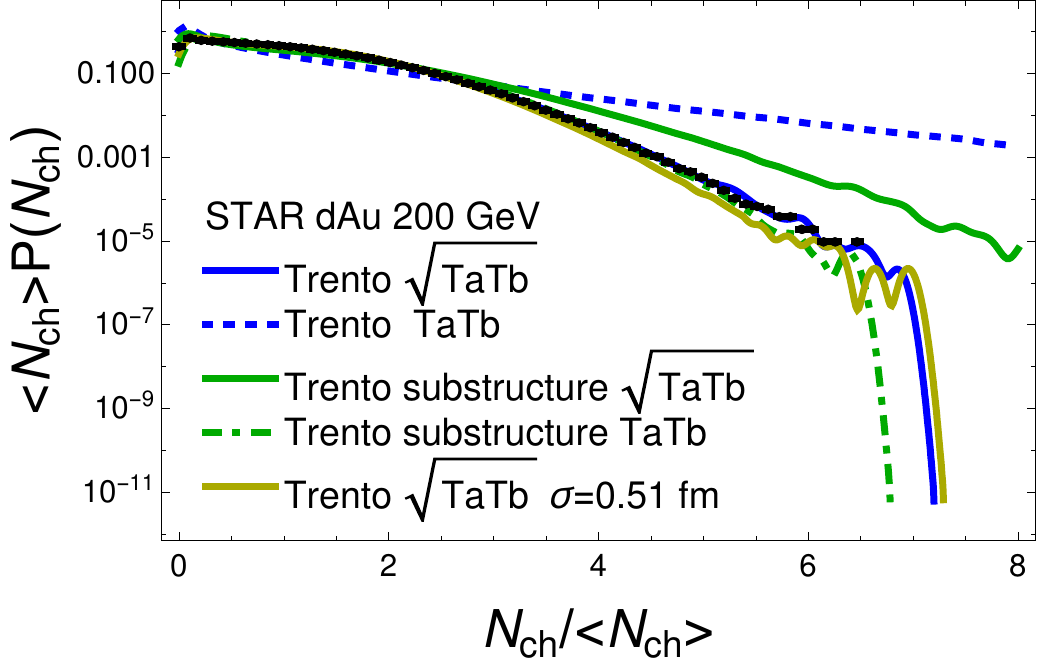}
	\caption{Multiplicity distributions in TRENTO for different parameter choices, compared to STAR data  \cite{Abelev:2008ab}.  For nucleons, unless otherwise noted we use a default width $\sigma = 0.3~\mathrm{fm}$.} 
	\label{f:dAu_Nch}
\end{figure}

\begin{figure}[!h]
	\includegraphics[width=\linewidth]{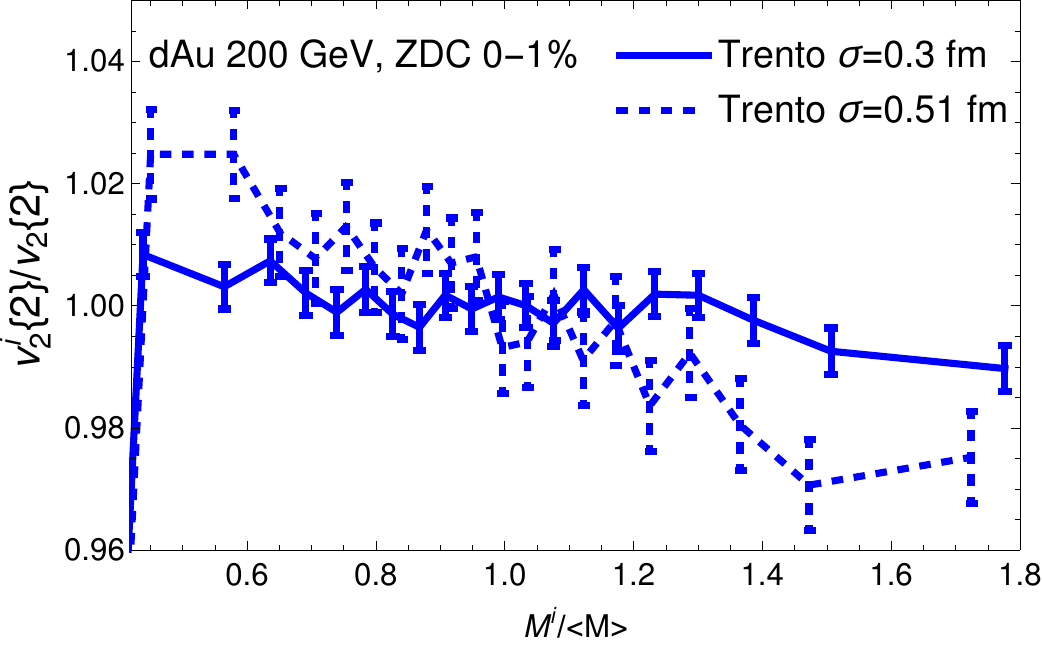}
	\caption{
		Dependence on the nucleon width parameter $\sigma$ in TRENTO for the $0-1\%$ most central $\XX{}{d}{}{Au}$ collisions by ZDC at $200~\mathrm{GeV}$.  Here we use $\sqrt{T_A T_B}$ multiplicity scaling.} 
	\label{f:ZDCdAusig}
\end{figure}

Another substantial effect of including nucleon substructure is on the distribution of multiplicity fluctuations, as shown in Fig.~\ref{f:dAu_Nch}.  Nucleon degrees of freedom with Gaussian width $\sigma = 0.3~\mathrm{fm}$ describe the STAR data perfectly, while the larger value $\sigma = 0.51~\mathrm{fm}$ slightly underpredicts the high-multiplicity tail.  This is consistent with the conclusions of Ref.~\cite{Giacalone:2017uqx} that a slight preference may exist for the smaller value $\sigma = 0.3~\mathrm{fm}$ in small systems.  As seen in Fig.~\ref{f:ZDCdAusig}, this change in the nucleon width parameter also leads to a small $\ord{2\%}$ change in the multiplicity dependence of $v_2 \{2\}$ in the initial conditions+hydrodynamics picture.  The multiplicity fluctuations including $n=6$ sub-nucleonic constituents for $\sqrt{T_A T_B}$ scaling also capture the trends of the experimental multiplicity distribution in Fig.~\ref{f:dAu_Nch} reasonably well, but overpredict the high-multiplicity tail.  This is not surprising, because the particular substructure parameters used here were tuned for $\XX{}{p}{208}{Pb}$ collisions at the LHC.  

\begin{figure}[t]
	\includegraphics[width=\linewidth]{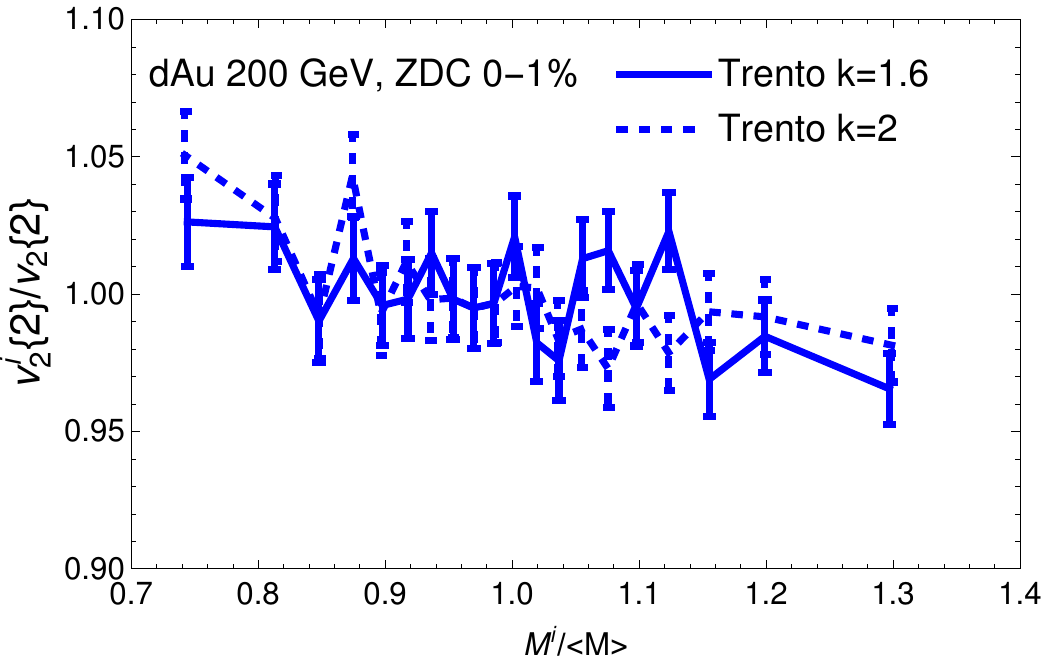}
	\caption{
		Dependence on the multiplicity fluctuation parameter $k$ in TRENTO for the $0-1\%$ most central $\XX{}{d}{}{Au}$ collisions by ZDC at $200~\mathrm{GeV}$.  Here we use $\sqrt{T_A T_B}$ multiplicity scaling.  (See also Fig.~\ref{f:dAu_Nch}).} 
	\label{f:dAu_k_vary}
\end{figure}

The multiplicity distribution for nucleons with linear $T_A T_B$ scaling fails to describe the experimental data in Fig.~\ref{f:dAu_Nch}, suggesting that a significant retuning of the model parameters may be needed for this new scaling choice.  It is also quite surprising that the multiplicity distribution for linear $T_A T_B$ scaling including nucleon substructure agrees so well with the data, given that this is a new configuration of TRENTO and no retuning has been performed.  We suspect this is simply an accident of the various competing changes and consider a thorough exploration of the parameter space to be necessary to fully understand the content of the model.  We leave such an analysis for future work, and we will primarily focus on the phenomenologically successful $\sqrt{T_A T_B}$ scaling from here on, unless otherwise noted.  We do note, however, that the most important quantity controlling the multiplicity fluctuations is the TRENTO parameter $k$, which will need to be re-fit for $\XX{}{d}{197}{Au}$ collisions at RHIC.  Thankfully, however, the flow harmonics of interest to us here seem to be relatively insensitive to corrections to the fluctuation parameter $k$ as shown in Fig.~\ref{f:dAu_k_vary}.  Thus the deviations in the multiplicity distribution seen in Fig.~\ref{f:dAu_Nch} may not strongly affect the resulting flow harmonics -- at least for nucleon degrees of freedom.

\begin{figure}[t]
	\includegraphics[width=\linewidth]{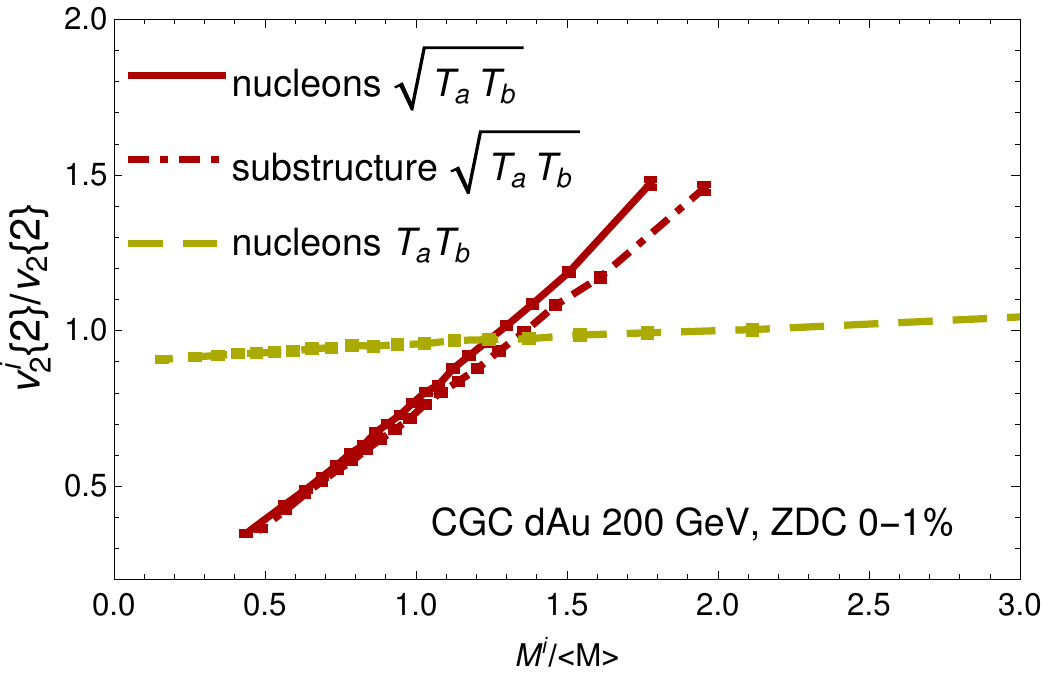}
	\caption{Sensitivity of the CGC picture to $T_A T_B$ vs. $\sqrt{T_A T_B}$ multiplicity scaling and to nucleonic substructure in $\XX{}{d}{197}{Au}$ collisions at RHIC.} 
	\label{f:CGC_TaTbscale}
\end{figure}

\begin{figure*}[t]
	\centering
	\begin{subfigure}[c]{0.49\linewidth}
		\includegraphics[width=\textwidth]
		{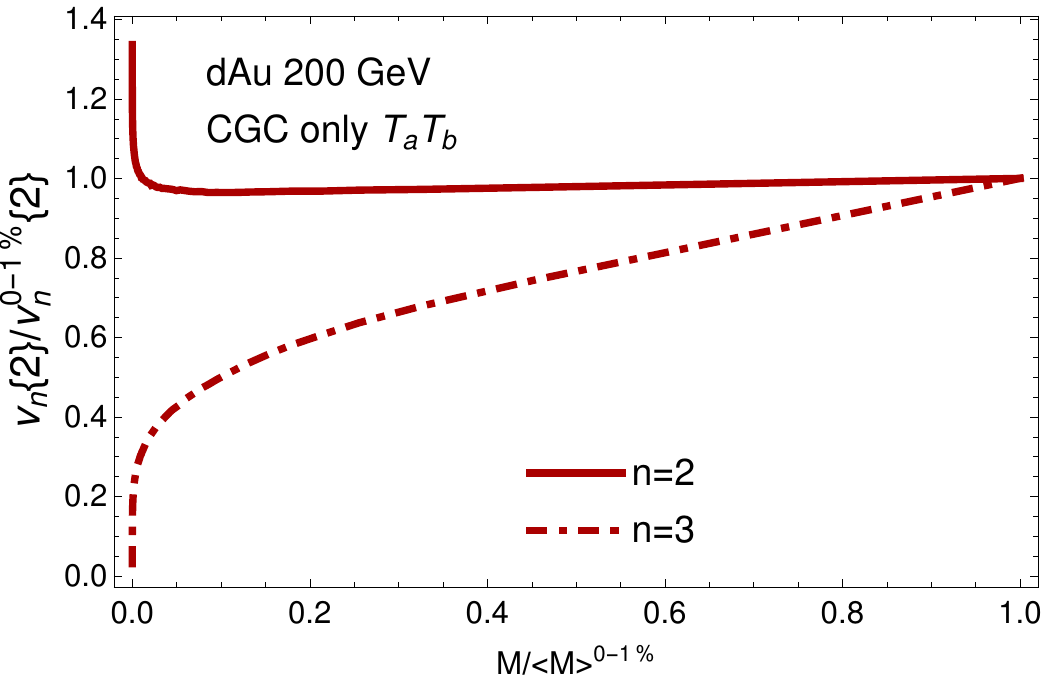}
	\end{subfigure}
	\begin{subfigure}[c]{0.49\linewidth}
		\includegraphics[width=\textwidth]
		{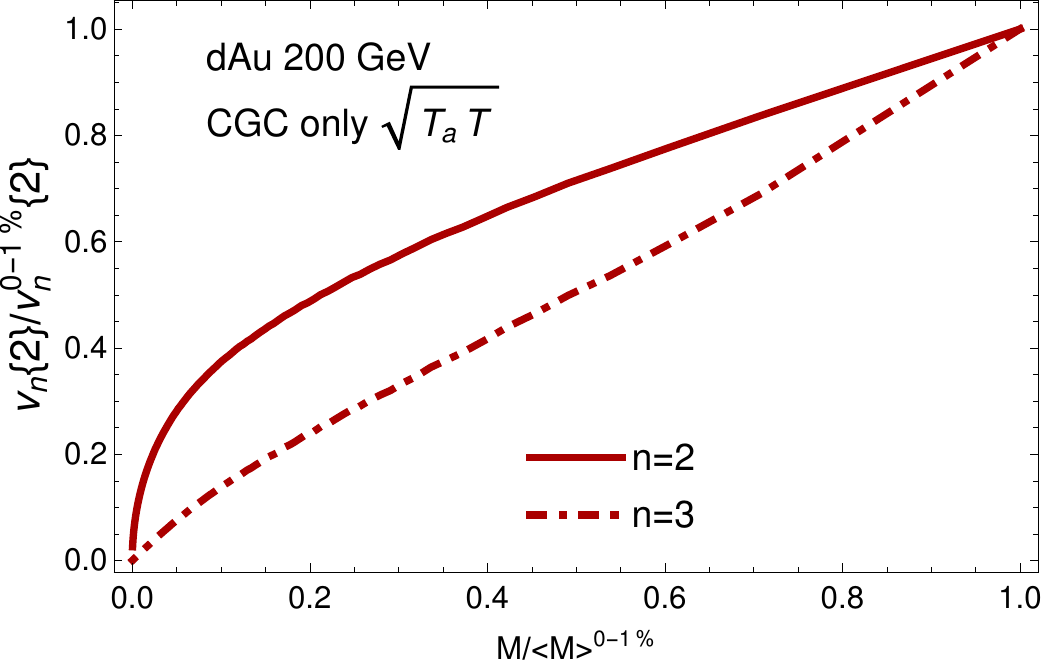}
	\end{subfigure}
	\caption{Elliptic and triangular flow versus multiplicity across the entire centrality range for $\XX{}{d}{}{Au}$ collisions at RHIC for the CGC mechanism \eqref{e:cum3v2}.  Here we compare the linear $T_A T_B$ multiplicity scaling (left panel) with $\sqrt{T_A T_B}$ scaling (right panel).  Note that the axes are scaled by the flow harmonics and multiplicity in the $0-1\%$ most central collisions, although results are shown for all centralities.}
	\label{f:CGCvn_v_mul}
\end{figure*}

\begin{figure}[h]
	\includegraphics[width=\linewidth]{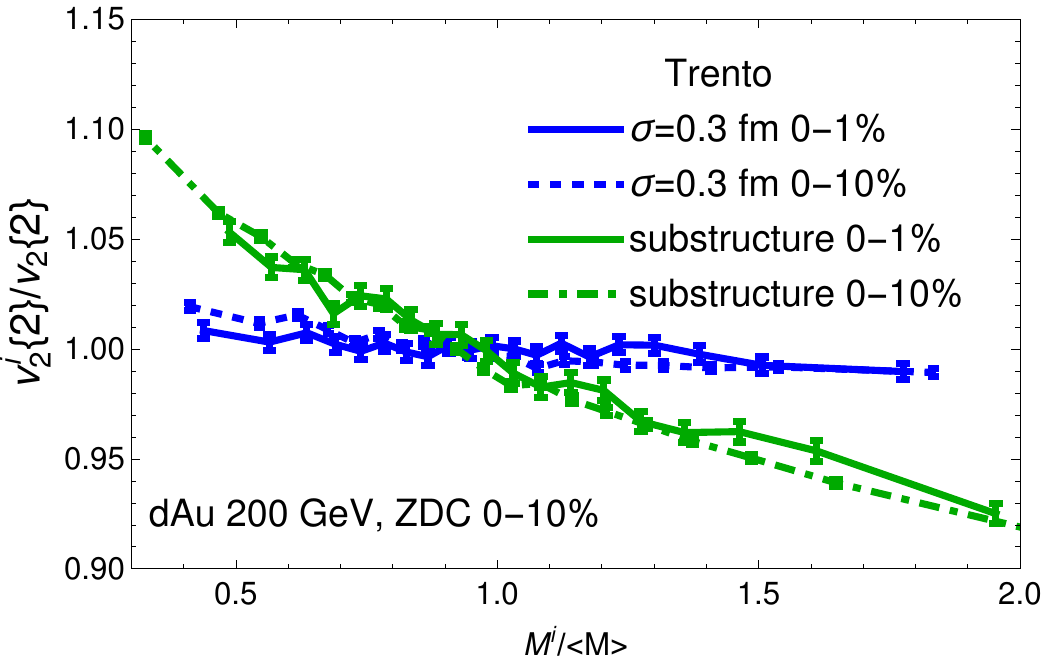}
	\caption{
		Comparison of elliptic flow in the central $0-10\%$ and ultracentral $0-1\%$ bins for $\XX{}{d}{197}{Au}$ collisions at RHIC in the hydrodynamics picture, for both nucleon and sub-nucleonic degrees of freedom.
	} 
	\label{f:dAu_centralitybins}
\end{figure}

The resulting sensitivity to the choice of multiplicity scaling and to the presence of nucleonic substructure is shown in  Fig.~\ref{f:CGC_TaTbscale} for the CGC picture.  As seen previously in Fig.~\ref{f:ZDCdAu}, there is a small flattening of the multiplicity dependence caused by the presence of nucleonic substructure for $\sqrt{T_A T_B}$ multiplicity scaling.  Far more significant is the dependence on the mechanism of multiplicity generation.  The smooth, deformed profile achieved with nucleon degrees of freedom and soft $\sqrt{T_A T_B}$ multiplicity scaling yields the greatest $\sim\ord{50\%}$ impact on the multiplicity dependence, and sharpening the multiplicity scaling to $T_A T_B$ flattens the multiplicity dependence tremendously.  Both of these effects can be interpreted as due to increasing the lumpiness of the density profiles, leading to a flattening of the multiplicity dependence.  We have also checked that this trend continues for the case of both linear $T_A T_B$ scaling and nucleonic substructure.

It is also interesting to observe the impact of the deformed deuteron geometry by comparing the $0-1\%$ ultracentral collisions seen in Fig.~\ref{f:ZDCdAu} against the same quantity over a wider centrality range.  A comparison of the ultracentral $0-1\%$ bin versus wider centrality selections is shown in Figs.~\ref{f:CGCvn_v_mul} and \ref{f:dAu_centralitybins} for the CGC and initial conditions+hydrodynamics pictures, respectively.  

For the CGC picture, we examine the full (minimum bias) centrality range in Fig.~\ref{f:CGCvn_v_mul} for both $T_A T_B$ and $\sqrt{T_A T_B}$ scaling.  The multiplicity dependence seen in the left panel of  Fig.~\ref{f:CGCvn_v_mul} for $T_A T_B$ scaling appears to reproduce beautifully the predictions $v_2 \{2\} \propto M^0$ and $v_3 \{2\} \propto M^{1/2}$ from Ref.~\cite{Mace:2018yvl}.  On the other hand, the corresponding multiplicity dependence that would arise from the $p=0$ scaling $\sqrt{T_A T_B}$ shown in the right panel of Fig.~\ref{f:CGCvn_v_mul} appears to have been significantly damped.  We study the multiplicity dependence of various collision systems and the impact of nuclear deformation in Appendix~\ref{app:CGCmult}.

For the initial conditions+hydrodynamics picture in Fig.~\ref{f:dAu_centralitybins}, the slopes and trends are practically identical for the two centrality cuts, both for nucleon degrees of freedom and for nucleonic substructure.  But in the case of sub-nucleonic fluctuations, the width of the multiplicity fluctuations in the $0-10\%$ bin has increased, leading to an increase in the overall magnitude of the substructure effect from $\ord{5\%}$ in the ultracentral bin to $\ord{10\%}$ in the central bin.  This suggests that selection cuts with wider bins that make it possible to capture wider multiplicity fluctuations with high statistics are a more promising approach to distinguishing the effects of sub-nucleonic fluctuations.

\subsection{$\XX{9}{Be}{197}{Au}$ and $\XX{9}{Be}{9}{Be}$ Collisions}
\label{ss:Be}

\begin{figure}[h]
	\includegraphics[width=\linewidth]{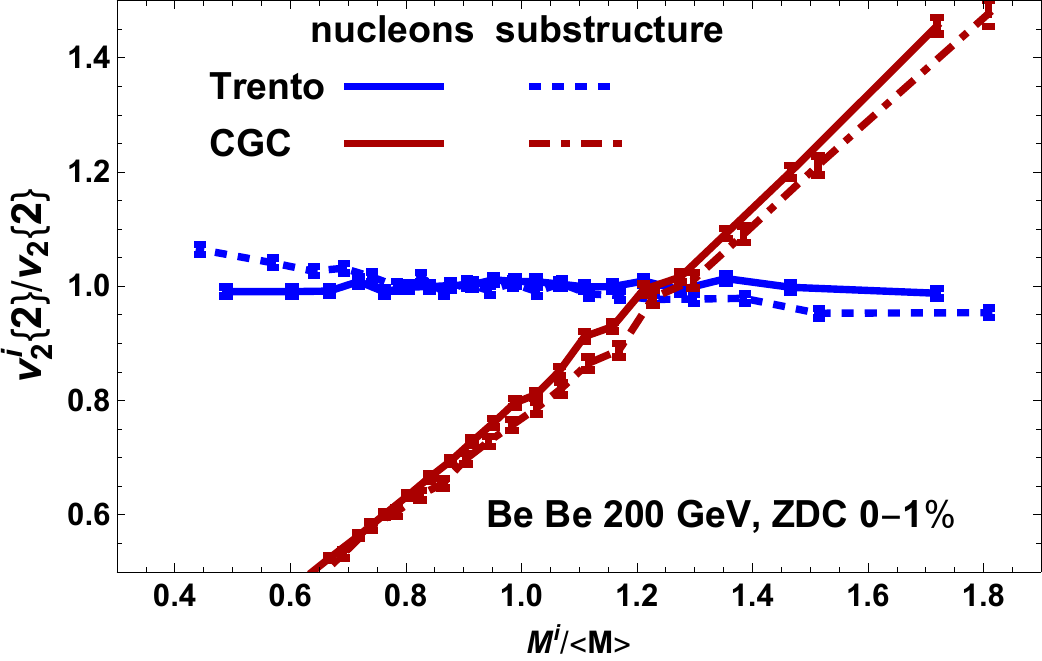}
	\caption{
		Elliptic flow versus multiplicity for the $0-1\%$ most central $\XX{9}{Be}{9}{Be}$ collisions by ZDC at $200~\mathrm{GeV}$ at RHIC, using $\sqrt{T_A T_B}$ multiplicity scaling.  Solid curves indicate nucleon degrees of freedom; dashed curves include $n=6$ sub-nucleonic constituents.
	} 
	\label{f:ZDCBe}
\end{figure}

\begin{figure}[h]
	\includegraphics[width=\linewidth]{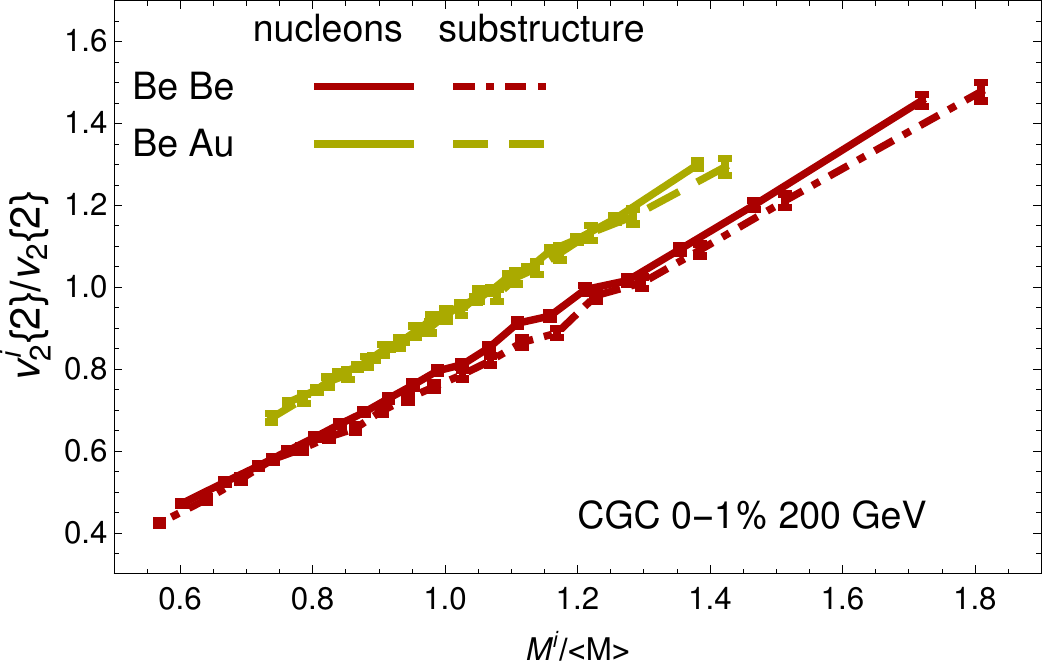}
	\caption{
		Sensitivity of the CGC picture to nucleonic substructure in ultracentral collisions of both $\XX{9}{Be}{9}{Be}$ and $\XX{9}{Be}{197}{Au}$ at RHIC.  Here we use $\sqrt{T_A T_B}$ multiplicity scaling.
	} 
	\label{f:BeCGC}
\end{figure}

\begin{figure}[h]
	\includegraphics[width=\linewidth]{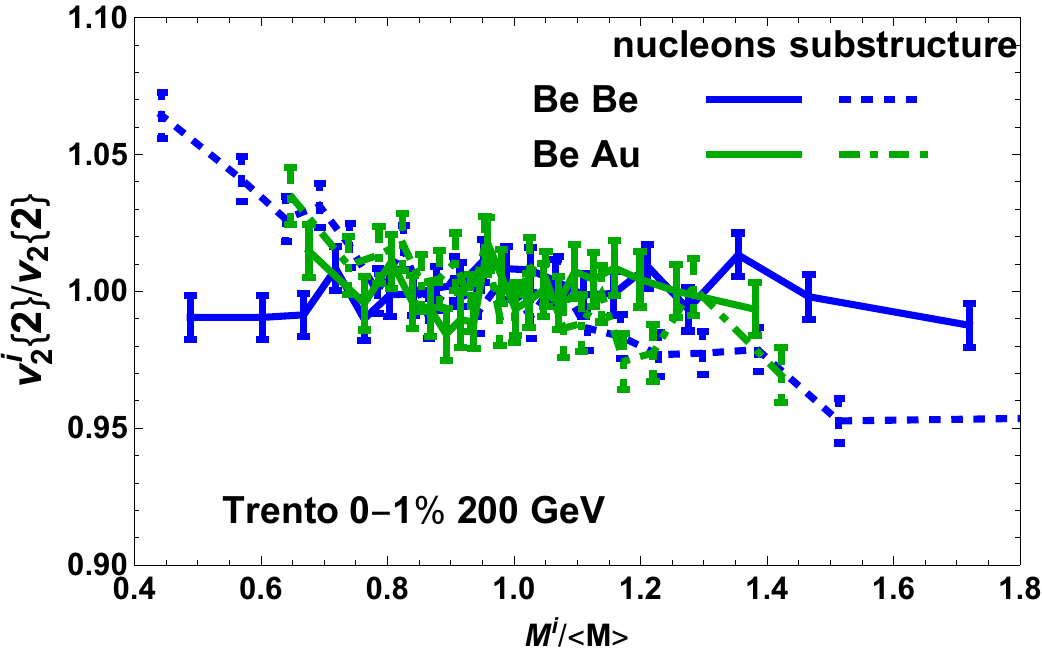}
	\caption{
		Sensitivity of the hydrodynamics picture to nucleonic substructure in ultracentral collisions of both $\XX{9}{Be}{9}{Be}$ and $\XX{9}{Be}{197}{Au}$ at RHIC.  Here we use $\sqrt{T_A T_B}$ multiplicity scaling.
	} 
	\label{f:BeAu}
\end{figure}

\begin{figure}[h]
	\includegraphics[width=\linewidth]{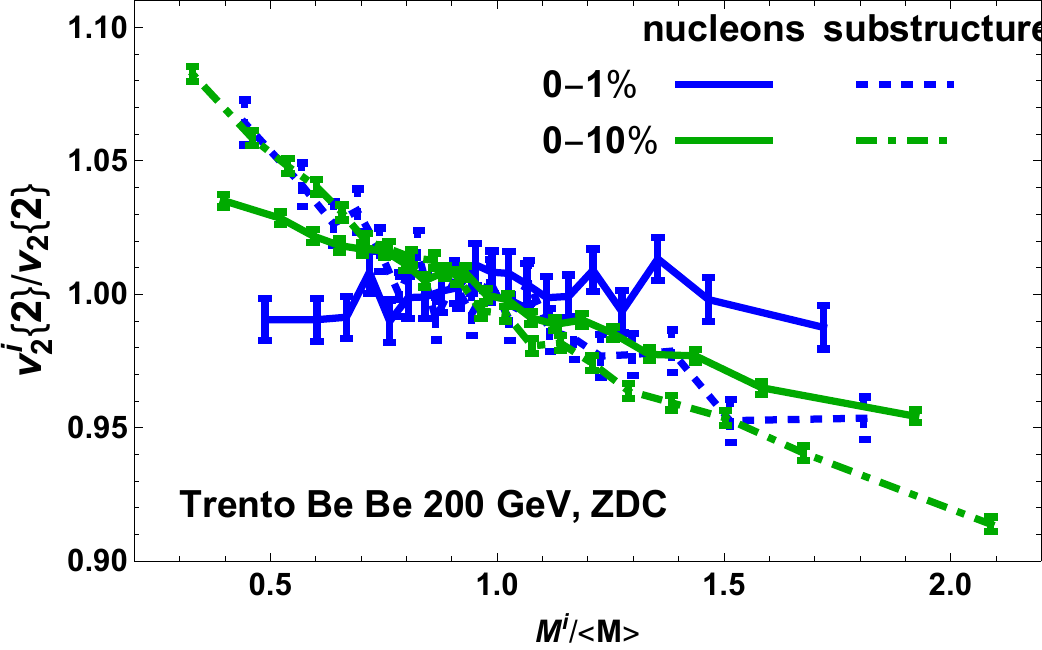}
	\caption{
		Comparison of elliptic flow in the central $0-10\%$ and ultracentral $0-1\%$ bins for $\XX{9}{Be}{9}{Be}$ collisions at RHIC in the hydrodynamics picture, for both nucleon and sub-nucleonic degrees of freedom.  Here we use $\sqrt{T_A T_B}$ multiplicity scaling.
	} 
	\label{f:Be010}
\end{figure}

A useful system intermediate to $\XX{}{d}{197}{Au}$ and $\XX{238}{U}{238}{U}$ collisions is $\XX{9}{Be}{197}{Au}$, since it collides the small, highly-deformed beryllium ion with the large spherical gold ion.  This highly asymmetric collision is to be contrasted with a symmetric collision of small deformed ions with each other, such as $\XX{9}{Be}{9}{Be}$.  Collisions involving ${}^9 \mathrm{Be}$ were previously proposed in Ref.~\cite{Bozek:2018xzy} as part of a program to study collisions of light polarized nuclei, although the polarization studies are not our main focus in this paper.  For $\XX{}{}{9}{Be}$ we use the parameters $R=1.4~\mathrm{fm}$, $a = 0.75~\mathrm{fm}$, $\beta_2 = 0.64$, $\beta_4 = 0.27$ from Ref.~\cite{Lukyanov:2013xea}; in comparison, the quadrupole deformation $\beta_2$ of  $\XX{238}{U}{}{}$ is only about $0.28$.  

An overview of the elliptic flow results for symmetric $\XX{9}{Be}{9}{Be}$ collisions in both the initial conditions+hydrodynamics and CGC pictures is shown in Fig.~\ref{f:ZDCBe}.  More detailed results comparing $\XX{9}{Be}{9}{Be}$ and $\XX{9}{Be}{197}{Au}$ collisions are shown in Fig.~\ref{f:BeCGC} for the CGC picture and in Fig.~\ref{f:BeAu} for the hydrodynamics picture.  The overall qualitative picture seen for $\XX{9}{Be}{9}{Be}$ collisions in Fig.~\ref{f:ZDCBe} is quite consistent with the results shown in Fig.~\ref{f:ZDCdAu} for $\XX{}{d}{197}{Au}$.  For nucleon degrees of freedom, the hydrodynamics picture produces a mostly constant $v_2 \{2\}$, with the addition of nucleonic substructure producing an $\ord{5 - 10\%}$ enhancement at low multiplicity and a small negative slope.  In contrast, the CGC mechanism increases almost linearly with multiplicity, with the addition of substructure leading to a slight decrease in the slope of the multiplicity dependence.

The elliptic flow vs. multiplicity results for the CGC picture with $\sqrt{T_A T_B}$ multiplicity scaling are shown in more detail in Fig.~\ref{f:BeCGC}, including both $\XX{9}{Be}{9}{Be}$ and $\XX{9}{Be}{197}{Au}$ collisions as well as for $\sigma=0.3~\mathrm{fm}$ nucleons and $n=6$ nucleonic substructure.  The results seen here are consistent with the ones shown in Figs.~\ref{f:ZDCdAu} and \ref{f:CGC_TaTbscale} for $\XX{}{d}{197}{Au}$: the effect of nucleonic substructure is to slightly flatten the multiplicity distribution due to the increasing lumpiness of the initial density profile.  We also note that the slope of the CGC effect is slightly greater in the collisions of $\XX{9}{Be}{}{}$ than it was for $\XX{}{d}{197}{Au}$: the same $\ord{60\%}$ effect on $v_2 \{2\}$ is achieved for $\XX{9}{Be}{}{}$ over a somewhat smaller range of multiplicity fluctuations than in $\XX{}{d}{197}{Au}$.  Presumably, this is due to the large deformation of $\XX{9}{Be}{}{}$.

Similar results for the initial conditions+hydrodynamics picture are shown in more detail in Fig.~\ref{f:BeAu}.  As with $\XX{}{d}{197}{Au}$ collisions shown in Figs.~\ref{f:ZDCdAu} and \ref{f:dAu_centralitybins}, for nucleon degrees of freedom the multiplicity dependence is quite flat, both for $\XX{9}{Be}{9}{Be}$ and $\XX{9}{Be}{197}{Au}$ collisions.  Turning on nucleonic substructure introduces a discernable negative slope to the multiplicity dependence, although the effects are hard to distinguish in the case of $\XX{9}{Be}{197}{Au}$ collisions.  But, while the slopes appear to be similar in both $\XX{9}{Be}{9}{Be}$ and $\XX{9}{Be}{197}{Au}$ collisions with sub-nucleonic fluctuations, the wider multiplicitiy fluctuations seen in the smaller $\XX{9}{Be}{9}{Be}$ system results in a more prominent effect.  

We can further enhance the width of the multiplicity fluctuations in $\XX{9}{Be}{9}{Be}$ collisions by relaxing our ultracentral $0-1\%$ cut to a $0-10\%$ centrality cut, as shown in Fig.~\ref{f:Be010}.  As expected, this does increase the width of the multiplicity fluctuations, increasing the magnitude of the signal of nucleonic substructure.  However, the $0-10\%$ cut also begins to capture a nontrivial ellipticity for nucleons; this generates a slope for the nucleon case and competes with the enhanced multiplicity fluctuations, making the ability to resolve nucleonic substructure nontrivial.  It would seem that the ultracentral $0-1\%$ events in $\XX{9}{Be}{9}{Be}$ collisions really select on round collision geometries, whereas a less extreme centrality cut reflects the elliptical shape of ${}^9 \mathrm{Be}$.

\begin{figure}[t]
	\includegraphics[width=\linewidth]{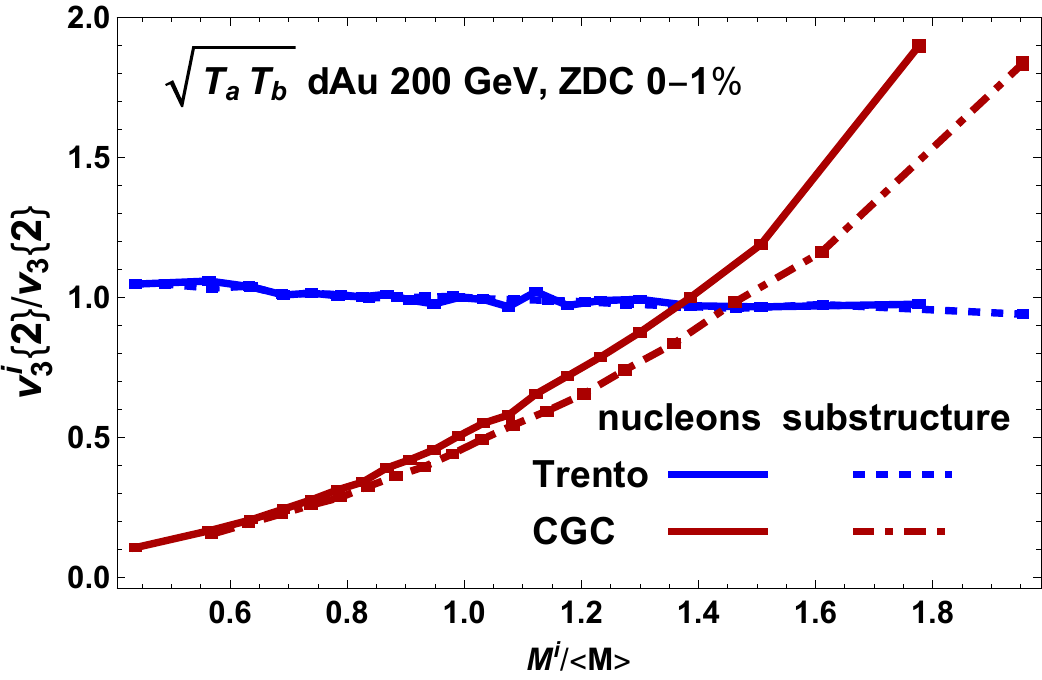}
	\caption{Triangular flow $v_3\{2\}$ in ultracentral $\mathrm{d} \mathrm{Au}$ collisions at RHIC for the hydrodynamics and CGC pictures, with and without $n=6$ sub-nucleonic constituents, using $\sqrt{T_A T_B}$ multiplicity scaling.} 
	\label{f:dAu_v3}
\end{figure}

\begin{figure}[t]
	\includegraphics[width=\linewidth]{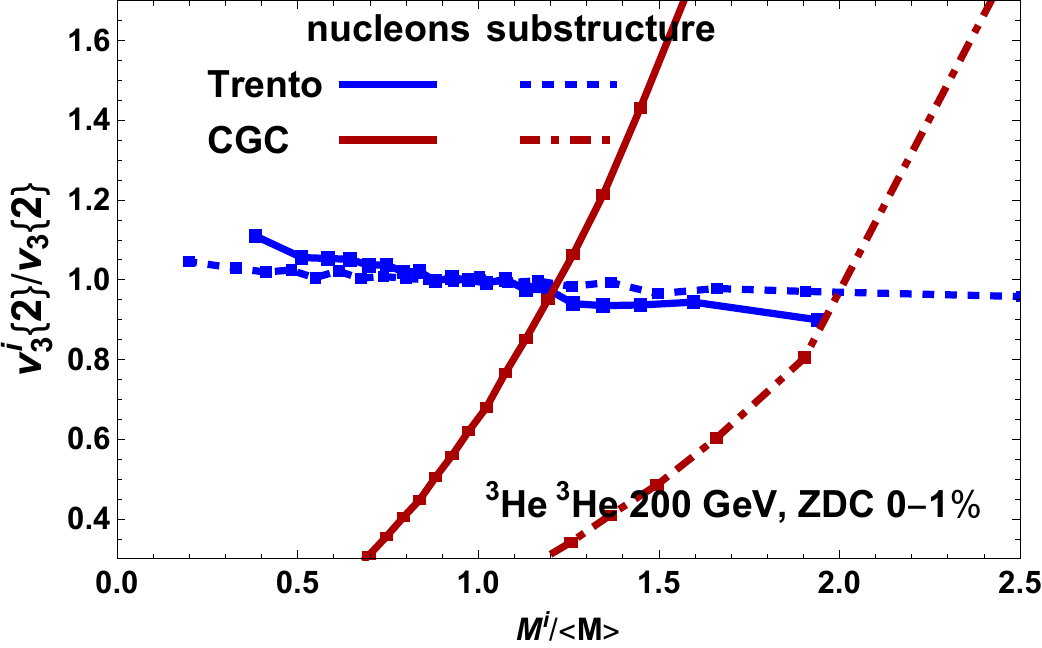}
	\caption{Triangular flow $v_3\{2\}$ in ultracentral $\XX{3}{He}{3}{He}$ collision at RHIC for the hydrodynamics and CGC pictures, with and without $n=6$ sub-nucleonic constituents, using $\sqrt{T_A T_B}$ multiplicity scaling.} 
	\label{f:ZDCHeHe}
\end{figure}

The emerging picture regarding nucleonic substructure is that, while the impact of substructure is fairly small, it can be more easily distinguished by enhancing the role of multiplicity fluctuations.  This can be achieved by considering ultracentral collisions of small symmetric systems such as $\XX{9}{Be}{9}{Be}$ rather than $\XX{9}{Be}{197}{Au}$.  Relaxing the ultracentral centrality cuts can also enhance the multiplicity fluctuations associated with nucleonic substructure.  This leads to a clear enhancement of the discriminating power between nucleon and sub-nucleonic degrees of freedom in Fig.~\ref{f:dAu_centralitybins} for $\XX{}{d}{197}{Au}$ collisions, but for larger deformed ions like ${}^9 \mathrm{Be}$ this multiplicity enhancement competes against the innate ellipticity of the nucleon distribution.  We expand upon this idea in greater detail for $\XX{3}{He}{}{Au}$ versus $\XX{3}{He}{3}{He}$ collisions in the following Section.  

\subsection{Triangular Flow and $\XX{3}{He}{3}{He}$ and $\XX{3}{He}{197}{Au}$ Collisions}
\label{ss:He}

Unlike the deformed ions that discussed previously, ${}^3 \mathrm{He}$ has an inherent triangular shape rather than elliptical shape.  Thus it is interesting to study the response of the triangular flow $v_3 \{2\}$ in collisions involving ${}^3 \mathrm{He}$, as compared to the other elliptically-deformed systems.  As a baseline to compare against the triangular shape of ${}^3 \mathrm{He}$, in Fig.~\ref{f:dAu_v3} we plot the triangular flow $v_3\{2\}$ for $\XX{}{d}{197}{Au}$ collisions for $\sqrt{T_A T_B}$ scaling, with and without nucleonic substructure.  As expected from the lack of a triangular geometry with nucleon degrees of freedom, the initial conditions+hydrodynamics response is almost completely flat.  Contrary to what might be expected, introducing nucleonic substructure does not appear to generate an inherent triangularity $\varepsilon_3$ from its fluctuating substructure -- at least not contributing to the ultracentral $0-1\%$ centrality bin.  The CGC picture, on the other hand, generates a triangular flow directly without needing an underlying triangular geometry $\varepsilon_3$, so it produces a positively-sloped curve both with and without substructure.  And, as with previous collision systems, turning on nucleonic substructure somewhat flattens the slope of the CGC mechanism.

In Fig.\ \ref{f:ZDCHeHe} we perform the same comparison for the hydrodynamics picture in $\XX{3}{He}{3}{He}$ collisions.  Because of the innate $\varepsilon_3$ of the triangular ${}^3 \mathrm{He}$ ion, the initial conditions+hydrodynamics picture possesses a negative slope even for nucleon degrees of freedom.  Turning on nucleonic substructure, however, {\it{flattens}} this behavior, smoothing out the multiplicity dependence associated with the triangularity of ${}^3 \mathrm{He}$.  The innate $\varepsilon_3$ at the level of nucleons leads to an  $\sim\ord{10\%}$ effect, whereas turning on substructure reduces this to an $\sim\ord{5\%}$ effect.  Interestingly, this substructure effect moves the triangularity of ${}^3 \mathrm{He}$ in an opposite direction from the ellipticity seen in ${}^9 \mathrm{Be}$ and the deuteron.  Meanwhile, the CGC picture shows a steeper slope for $v_3 \{2\}$ than for $v_2 \{2\}$ in accordance with the dependence on a higher power $T_A^3 T_B^3$ of the nuclear profiles in Eq.~\eqref{e:cum2}.  As seen in the other collision systems previously, turning on nucleonic substructure dilutes the effect and flattens the slope somewhat.  

\begin{figure}[t]
	\includegraphics[width=\linewidth]{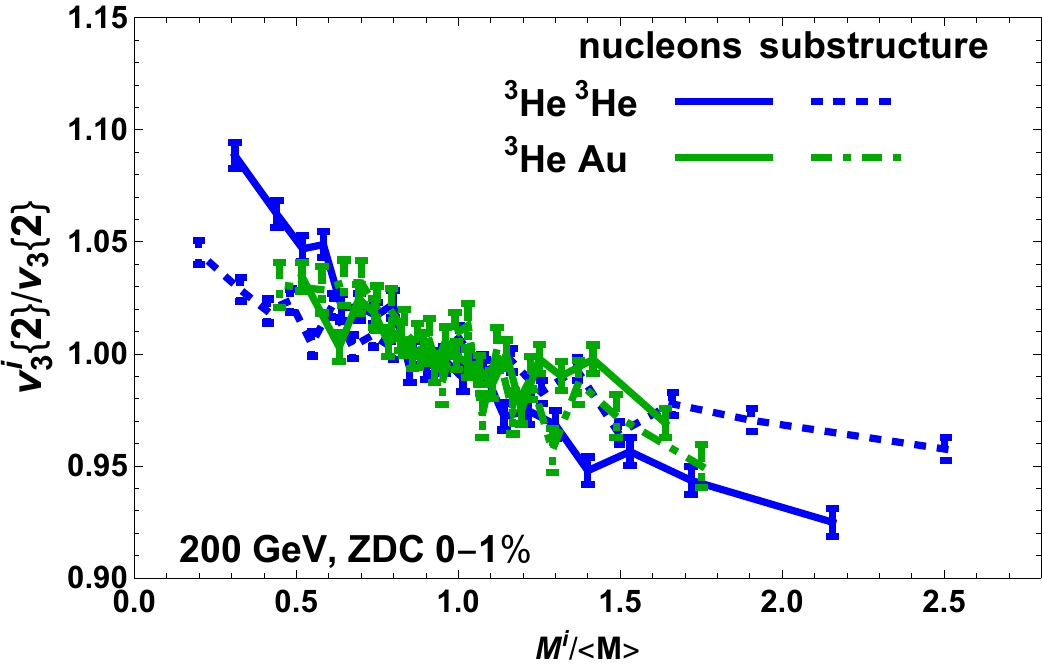}
	\caption{
		Comparison of triangular flow $v_3\{2\}$ in ultracentral $\XX{3}{He}{3}{He}$ versus $\XX{3}{He}{197}{Au}$ collisions
		collision at RHIC for the hydrodynamics picture, with and without $n=6$ sub-nucleonic constituents, using $\sqrt{T_A T_B}$ multiplicity scaling.} 
	\label{f:ZDCHe3Au}
\end{figure}

Previously, we saw that the substructure effect in $\XX{9}{Be}{9}{Be}$ versus $\XX{9}{Be}{197}{Au}$ collisions in Fig.~\ref{f:BeAu} led to comparable slopes, with the distinguishing feature being the wider multiplicity fluctuations present in the smaller collision system.  This difference enhanced an effect which may have been unidentifiable in $\XX{9}{Be}{197}{Au}$ collisions to be an $\ord{5\%}$ effect in $\XX{9}{Be}{9}{Be}$ collisions.  Accordingly, we expect that symmetric collisions of small deformed ions will be more discriminating (within a hydrodynamics picture) to the presence or absence of sub-nucleonic fluctuations.  We can further test this interpretation by extending it to the $v_3 \{2\}$ sector and comparing $\XX{3}{He}{3}{He}$ versus $\XX{3}{He}{197}{Au}$ collisions, which are shown in Fig.\ \ref{f:ZDCHe3Au}.  As anticipated, we find in Fig.\ \ref{f:ZDCHe3Au} that $\XX{3}{He}{3}{He}$ collisions are more sensitive than $\XX{3}{He}{197}{Au}$ collisions to the presence of sub-nucleonic fluctuations.  The reason for this discriminating power for $v_3 \{2\}$ in ${}^3 \mathrm{He}$ is opposite to the reason for its discriminating power for $v_2 \{2\}$ in $\XX{}{d}{197}{Au}$ and ${}^9 \mathrm{Be}$ -- a suppression due to substructure, rather than an enhancement -- but it is a discriminator nonetheless.

\begin{figure}[t]
	\includegraphics[width=\linewidth]{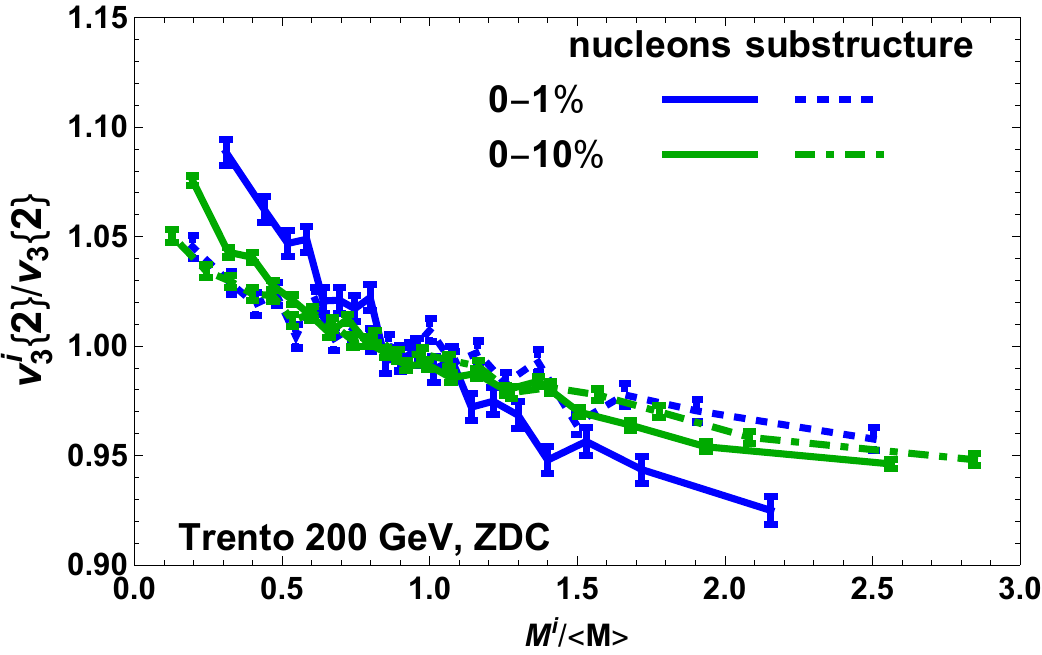}
	\caption{
		Comparison of elliptic flow in the central $0-10\%$ and ultracentral $0-1\%$ bins for $\XX{3}{He}{3}{He}$ collisions at RHIC in the hydrodynamics picture, for both nucleon and sub-nucleonic degrees of freedom.  Here we use $\sqrt{T_A T_B}$ multiplicity scaling.} 
	\label{f:He3010}
\end{figure}

Similarly, we saw in Fig.~\ref{f:dAu_centralitybins} that the increase in multiplicity fluctuations in going from $0-1\%$ ultracentral collisions to $0-10\%$ central collisions for $\XX{}{d}{197}{Au}$ magnified an $\ord{5\%}$ substructure effect to an $\ord{10\%}$ effect.  However, we also saw in Fig.~\ref{f:Be010} that for $\XX{9}{Be}{9}{Be}$ collisions this enhancement of the substructure signal due to increased multiplicity was partially offset by an increased elliptic flow for nucleon degrees of freedom only.  For $\XX{3}{He}{3}{He}$ collisions binned with $0-1\%$ versus $0-10\%$ centrality cuts shown in Fig.~\ref{f:He3010}, the story is similar.  Keeping in mind that nucleonic substructure in this case leads to a {\it{suppression}} of $v_3 \{2\}$, we see that for $0-1\%$ ultracentral collisions, the separation between nucleonic and sub-nucleonic degrees of freedom is around $\ord{5\%}$.  Relaxing the centrality cuts to $0-10\%$ central events does somewhat increase the range of multiplicity fluctuations for nucleonic substructure, but it also leads to a simultaneous reduction in the $v_3 \{2\}$ for nucleons.  This again reduces the discriminating power between nucleon and sub-nucleonic degrees of freedom in the looser $0-10\%$ centrality cuts.

\section{Conclusions}
\label{Sec:Conclusions}

In this paper we studied ultracentral collisions of deformed ions, specifically studying the scaling of elliptic flow $v_2\{2\}$ and triangular flow $v_3\{2\}$ versus multiplicity.  This work was motivated in part by the tension between the prediction of  Ref.~\cite{Kovchegov:2013ewa} that gluon correlations in the CGC should exhibit a positive correlation of $v_2\{2\}$ with multiplicity in central $\XX{238}{U}{238}{U}$ collisions, and the anticorrelation seen in STAR data.  We therefore compared these opposing correlations of elliptic flow versus multiplicity across a multitude of smaller ultracentral collisional systems where initial-state CGC effects may play a more significant role: $\XX{238}{U}{238}{U}$, $\XX{}{d}{197}{Au}$, $\XX{9}{Be}{9}{Be}$, $\XX{9}{Be}{197}{Au}$, $\XX{3}{He}{3}{He}$, and $\XX{3}{He}{197}{Au}$.  Across all systems, we find that this qualitative distinction holds, with the CGC picture generally leading to a positive correlation between elliptic flow and multiplicity and the initial conditions+hydrodynamics picture leading to an anticorrelation.  A comparable picture also holds for the triangular flow arising in $\XX{}{d}{197}{Au}$, $\XX{3}{He}{3}{He}$, and $\XX{3}{He}{197}{Au}$ collisions.

We also study the dependence of these results on a number of parameters and model choices.  These include computing the initial conditions from nucleon versus nucleonic structure, computing the multiplicity from a given event proportional to $\sqrt{T_A T_B}$ versus $T_A T_B$, and varying the nucleon width.  The size of the nucleon width makes a small $\ord{2\%}$ difference in the slope for the initial conditions+hydrodynamics picture (see Fig.~\ref{f:ZDCdAusig}), with variations in the multiplicity fluctuation parameter $k$ having a smaller effect shown in Fig.~\ref{f:dAu_k_vary}, within error bars.  

The inclusion of $n=6$ sub-nucleonic constituents within TRENTO v2.0 leads to a systematic effect which is opposite in its enhancement of ellipticity in the deuteron and in ${}^9 \mathrm{Be}$ and its suppression of triangularity in ${}^3 \mathrm{He}$.  These effects are difficult to distinguish in asymmetric collisions such as $\XX{9}{Be}{197}{Au}$ and $\XX{3}{He}{197}{Au}$, but can be enhanced by increasing the multiplicity fluctuations using small symmetric collisions such as $\XX{9}{Be}{9}{Be}$ or $\XX{3}{He}{3}{He}$.  Using a looser $0-10\%$ centrality cut does significantly enhance the substructure signal in $\XX{}{d}{197}{Au}$ collisions, although the discriminating power in this wider bin is reduced for somewhat larger ions.  Taken together, these results confirm our interpretation of central and ultracentral collisions involving small deformed ions as being sensitive to sub-nucleonic fluctuations.  This holds true both in for elliptic and triangular flow, across a range of collision systems.

The CGC picture generally obtains positive correlations between the flow harmonics and multiplicity in deformed systems, with the slope of the correlation being the greatest for small asymmetric collisions with nucleon degrees of freedom and $\sqrt{T_A T_B}$ multiplicity scaling, which produces fairly smooth density profiles.  Increasing the lumpiness of the density profile, either by turning on nucleon substructure or changing to the sharper $T_A T_B$ multiplicity scaling, tends to flatten out the distribution.

In this paper we have focused exclusively on measurements for various ions at RHIC, based on the successful use of the STAR ZDC to select on tip-on-tip versus side-on-side collision geometries of ${}^{238} \mathrm{U}$.  One further possibility for future study using the unique capabilities of RHIC would include the interplay between polarization and deformation for light nuclei like ${}^9 \mathrm{Be}$; as is well-known in the spin community, polarizing a hadron can induce a corresponding spatial deformation of its internal constituents \cite{Burkardt:2002hr}.  The analysis performed here can also in principle be extended to the LHC, if a comparable selection on ultracentral events by spectator binning in the ZDC is possible; we leave this possibility for future work.

\section*{Acknowledgments}
The authors wish to thank 
K. Dusling,
T. Lappi, 
M. Luzum,
M. Mace, 
S. Moreland,
J. Nagle, 
B. Schenke, 
S. Schlichting, 
P. Tribedy,
R. Venugopalan, and
W. Zajc
for useful discussions.
The authors acknowledge support from the US-DOE Nuclear Science Grant No. DE-SC0019175, the Alfred P. Sloan Foundation, and the Office of Advanced Research Computing (OARC) at Rutgers, The State University of New Jersey for providing access to the Amarel cluster and associated research computing resources that have contributed to the results reported here.  D.E.W. acknowledges support from the Zuckerman STEM Leadership Program.  M.D.S. and D.E.W. wish to thank the Department of Energy's Institute for Nuclear Theory at the University of Washington for its hospitality while facilitating a portion of this work.

\appendix

\section{Centrality binning method}
\label{sec:center}

Here we explore which centrality binning method is closest to experimental data.  The simplest way to understand what STAR's ZDC measures is it essentially the number of nucleons that {\it{don't}} participate in the collision.  This would be roughly equivalent to $N_{ZDC}=2A-N_{\mathrm{part}}$ and for ultracentral collisions $N_{ZDC}\rightarrow 0$.  From this standpoint, we anticipate that binning by $N_{\mathrm{part}}$ would then be the most effective way to select on the $0-1\%$ most central events.  However, after that initial selection is made the events are binned by multiplicity.  Thus, we conclude that one first must bin by $N_{\mathrm{part}}$ to obtain the $0-1\%$ centrality class but then one must bin by multiplicity, which from an initial conditions standpoint corresponds to total entropy.  (Note that here we use $M\sim \frac{S_0}{4}$.)  

\begin{figure}[!h]
	\includegraphics[width=\linewidth]{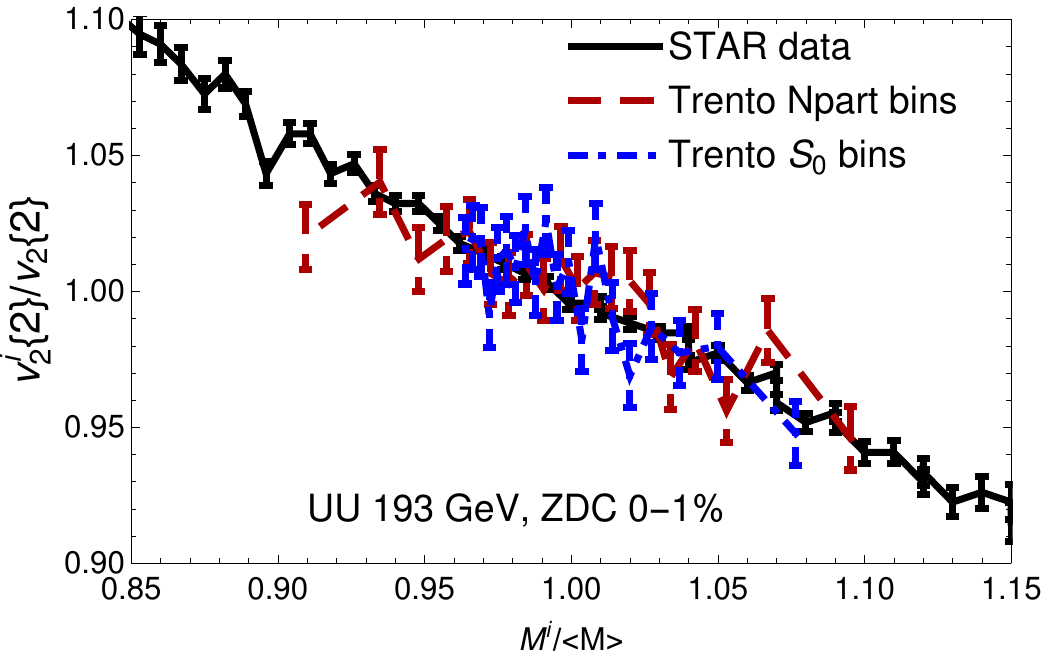}
	\caption{$UU$ collision events in the $0-1\%$ most central collisions (selecting on $N_{\mathrm{part}}$ vs. entropy $S_0$) where the average multiplicity is $\langle M\rangle$ and the flow is $v_n\{2\}$ for the entire centrality class.  20 subdivisions $i$ in multiplicity are made and their corresponding $v_n^i \{2\}$ is calculated. } 
	\label{f:ZDCbin}
\end{figure}

In order to demonstrate this we compare binning the $0-1\%$ by either $N_{\mathrm{part}}$ or $S_0$ as shown in Fig.\ \ref{f:ZDCbin}.  One notices immediately that binning by $N_{\mathrm{part}}$ provides a much wider range in multiplicity compared to binning by $S_0$ such that $N_{\mathrm{part}}$ binning is more consistent with experimental data.  Certainly, experimental data has even a wider range in multiplicity.  This may be simply due to statistics (for this result we ran 3 million events over all centralities, which is then 30,000 in the $0-1\%$ bin) but also may be attributed to differences between theory and the detector (e.g the detector can only measure charged particles whereas the multiplicity here includes all hadrons).  

In Fig.\ \ref{f:ZDCbin} we only show the results for TRENTO but we note that the CGC case has equivalent results (just with the inverse slope).

\section{Deformation Parameters for for ${}^{238} \mathrm{U}$}
\label{sec:Def}

Another natural question is if these symmetric ultracentral collisions can be used to constrain nuclear deformations. In order to investigate this, we use two different parameters sets for ${}^{238} \mathrm{U}$ which have been used in recent papers \cite{Giacalone:2018apa,Schenke:2019ruo}.  We do note, however, that in this paper we use $a=0.55$ rather than $a=0.6$ for the configuration in Ref.~\cite{Giacalone:2018apa} due to comparisons to data across all centralities with TRENTO+v-USPhydro calculations, which will be shown in an upcoming paper.  

\begin{figure}[!h]
	\includegraphics[width=\linewidth]{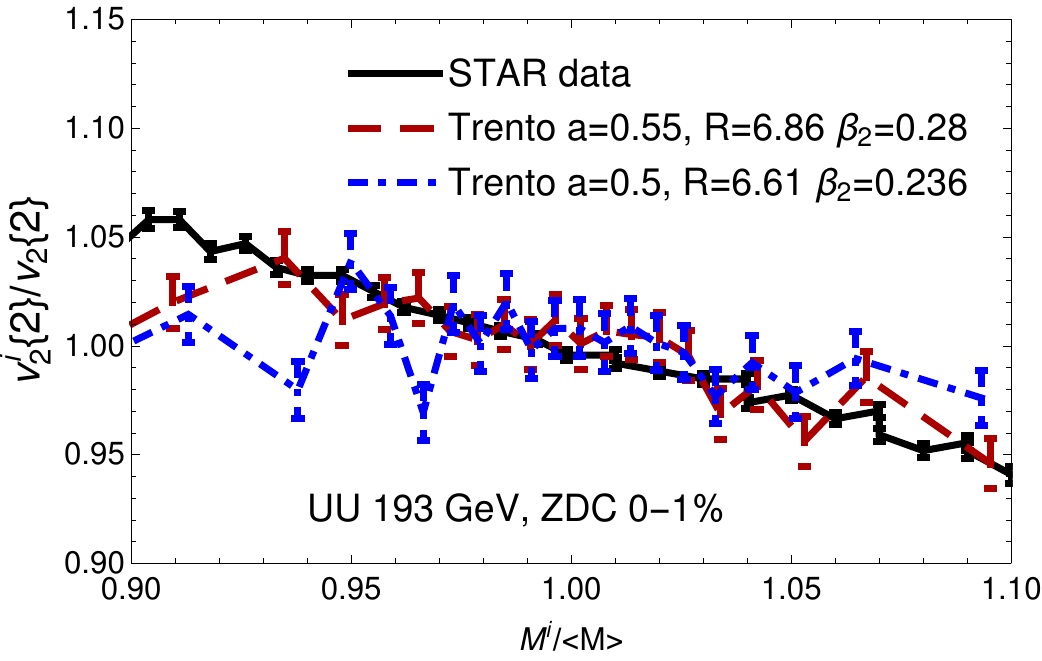}\\
	\includegraphics[width=\linewidth]{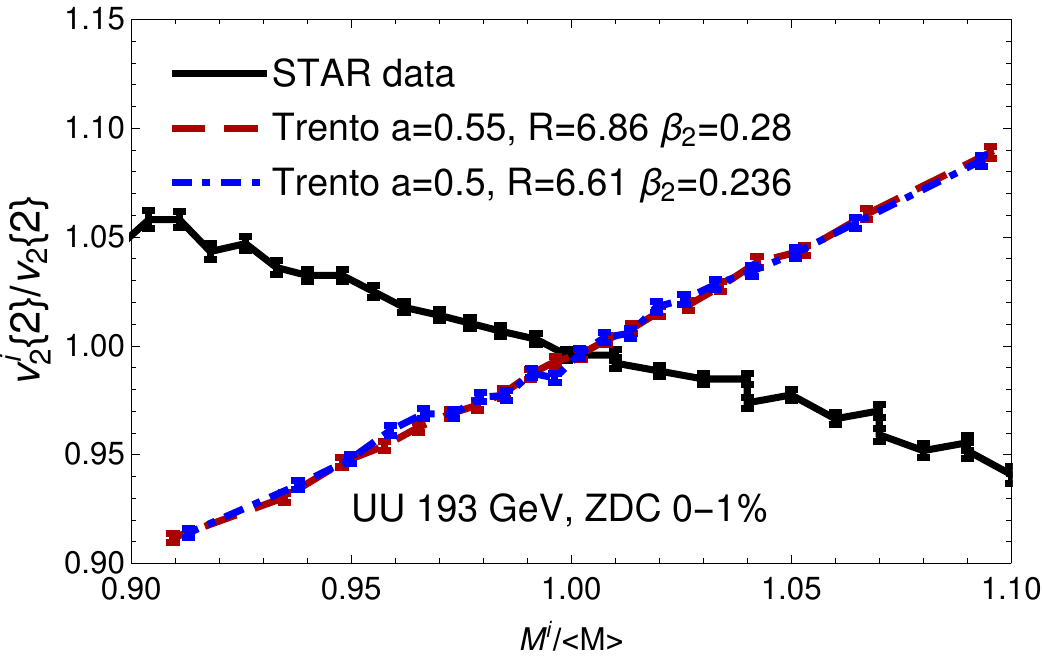}
	\caption{$\XX{238}{U}{238}{U}$ collision events in the $0-1\%$ most central collisions (selecting on $N_{\mathrm{part}}$) where the average multiplicity is $\langle M\rangle$ and the flow is $v_n\{2\}$ for the entire centrality class.  20 subdivisions $i$ in multiplicity are made and their corresponding $v_n^i \{2\}$ is calculated.  Comparisons between the initial conditions+hydrodynamics picture (top panel) and CGC picture (bottom panel) are shown. } 
	\label{f:ZDCUUdef}
\end{figure}

In Fig.\ \ref{f:ZDCUUdef} we compare the two different parameter sets for ${}^{238} \mathrm{U}$ and find that there is a slight preference for $a=0.55$, $R=6.86$, and $\beta_2=0.28$ in these comparisons, although we note that this should be taken in conjunction with other flow observable comparisons, which will be shown in future work.  Overall, we conclude that the deformation parameters play a small role in this observable but other flow observables may be more useful for constraining Wood-Saxon parameters.  For the CGC picture we see no difference whatsoever in Fig.\ \ref{f:ZDCUUdef} when we vary the Wood-Saxon parameters, since the CGC calculation is sensitive only to the multiplicity and not to the geometry itself.

\section{Multiplicity Dependence in the CGC}
\label{app:CGCmult}

In this Appendix we examine the simple theoretical expressions for the multiplicity dependence of the flow harmonics in the CGC picture and the role of nuclear deformations and gradients in modifying them.

The flow harmonics of two-gluon correlations arising in the CGC are expressed in Eq.~\eqref{e:cum2}, which we can rewrite as
\begin{subequations}
	\begin{align}
	v_2 \{2\}	&\sim
	\sqrt{\frac{
			\int d^2 x_\bot \, T_A^2 (\vec{x}_\bot) \, T_B^2 (\vec{x}_\bot)		
		}
		{
			\left[ \int d^2 x_\bot \frac{dN}{d^2 x} \right]^2
	}} 
	\\
	v_3 \{2\}	&\sim
	\sqrt{\frac{
			\int d^2 x_\bot \, T_A^3 (\vec{x}_\bot) \, T_B^3 (\vec{x}_\bot)		
		}
		{
			\left[ \int d^2 x_\bot \frac{dN}{d^2 x} \right]^2
	}} ,
	\end{align}
\end{subequations}
where we have omitted various proportionality factors to focus on the multiplicity dependence.  The translation from a dependence on $T_A , T_B$ to a multiplicity dependence varies based on how the multiplicity is calculated.  The multiplicity dependence one naturally obtains in the CGC uses a linear $T_A T_B$ scaling \eqref{e:TaTb}, but the most phenomenologically successful multiplicity dependence implemented in TRENTO uses a softer $\sqrt{T_A T_B}$ scaling \eqref{e:sqrtTaTb}.  Depending on which framework is used, therefore, we have one of the following expressions for the multiplicity density arising from the collision profile:
\begin{subequations} 
	\begin{align}	\label{e:Tlinear}
	\frac{dN}{d^2 x} &\sim T_A (\vec{x}_\bot) \, T_B (\vec{x}_\bot) ,
	\\	\label{e:Tsqrt}
	\frac{dN}{d^2 x} &\sim \sqrt{T_A (\vec{x}_\bot) \, T_B (\vec{x}_\bot)} .
	\end{align}
\end{subequations}
In terms of multiplicity alone, this then gives for the flow harmonics
\begin{subequations} \label{e:CGCmult}
	\begin{align}	\label{e:CGCmult1}
	v_2 \{2\} &\overset{T_A T_B}{\sim}
	\sqrt{\frac{
			\int d^2 x_\bot \, \left( \frac{dN}{d^2 x} \right)^2
		}
		{
			\left[ \int d^2 x_\bot \frac{dN}{d^2 x} \right]^2
	}} 
	&\sim& \qquad \mathrm{const}
	\\	\label{e:CGCmult2}
	v_3 \{2\} &\overset{T_A T_B}{\sim}
	\sqrt{\frac{
			\int d^2 x_\bot \, \left( \frac{dN}{d^2 x} \right)^3
		}
		{
			\left[ \int d^2 x_\bot \frac{dN}{d^2 x} \right]^2
	}} 
	&\sim& \qquad \sqrt{\frac{dN}{d^2 x}}
	\\ \notag \\	\label{e:CGCmult3}
	v_2 \{2\} &\overset{\sqrt{T_A T_B}}{\sim}
	\sqrt{\frac{
			\int d^2 x_\bot \, \left( \frac{dN}{d^2 x} \right)^4
		}
		{
			\left[ \int d^2 x_\bot \frac{dN}{d^2 x} \right]^2
	}} 
	&\sim& \qquad \frac{dN}{d^2 x}
	\\		\label{e:CGCmult4}
	v_3 \{2\} &\overset{\sqrt{T_A T_B}}{\sim}
	\sqrt{\frac{
			\int d^2 x_\bot \, \left( \frac{dN}{d^2 x} \right)^6
		}
		{
			\left[ \int d^2 x_\bot \frac{dN}{d^2 x} \right]^2
	}} 
	&\sim& \qquad \left(\frac{dN}{d^2 x}\right)^2 ,
	\end{align}
\end{subequations}
where in each case the last scaling on the right-hand side follows if the multiplicity dependence is roughly independent of the spatial geometry, $\frac{dN}{d^2 x} \approx \mathrm{const}$.  This approximate scaling can be satisfied if the collision region is not too large compared to the spatial gradients of the multiplicity density in a given framework.

This analysis is essentially the basis of the back-of-the envelope estimate by Mace et al. \cite{Mace:2018yvl}, concluding that $v_2 \{2\} \propto (N_{\mathrm{ch}})^0$ and $v_3 \{2\} \propto (N_{\mathrm{ch}})^{1/2}$.  In Ref.~\cite{Mace:2018yvl}, the authors attribute multiplicity fluctuations to the fluctuations of the dilute projectile in (semi)dilute/dense collisions such as $\XX{}{d}{}{Au}$.  Thus, although their calculation differs from the (semi)dilute / (semi)dilute calculation presented here, the two agree on the multiplicity of $v_n \{2\}$ dependence through the dominance of the (semi)dilute projectile.  The back-of-the-envelope estimates in Ref.~\cite{Mace:2018yvl} focus on the role of color-charge fluctuations in generating multiplicity fluctuations, without explicitly factoring in the geometry dependence of these fluctuations (although they do include event by event geometry fluctuations in their calculation).  As seen clearly in Eqs.~\eqref{e:CGCmult}, this estimate corresponds to a gradient expansion in which the multiplicity density $\frac{dN}{d^2 x}$ is slowly varying.

\begin{figure*}
	\centering
	\begin{subfigure}[c]{0.18\linewidth}
		\includegraphics[width=\textwidth]
		{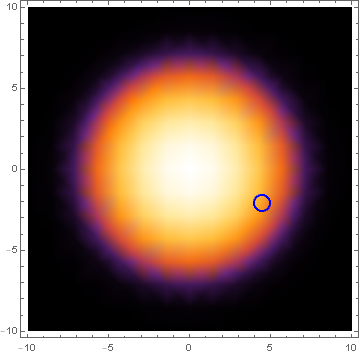}
	\end{subfigure}
	\begin{subfigure}[c]{0.4\linewidth}
		\includegraphics[width=\textwidth]
		{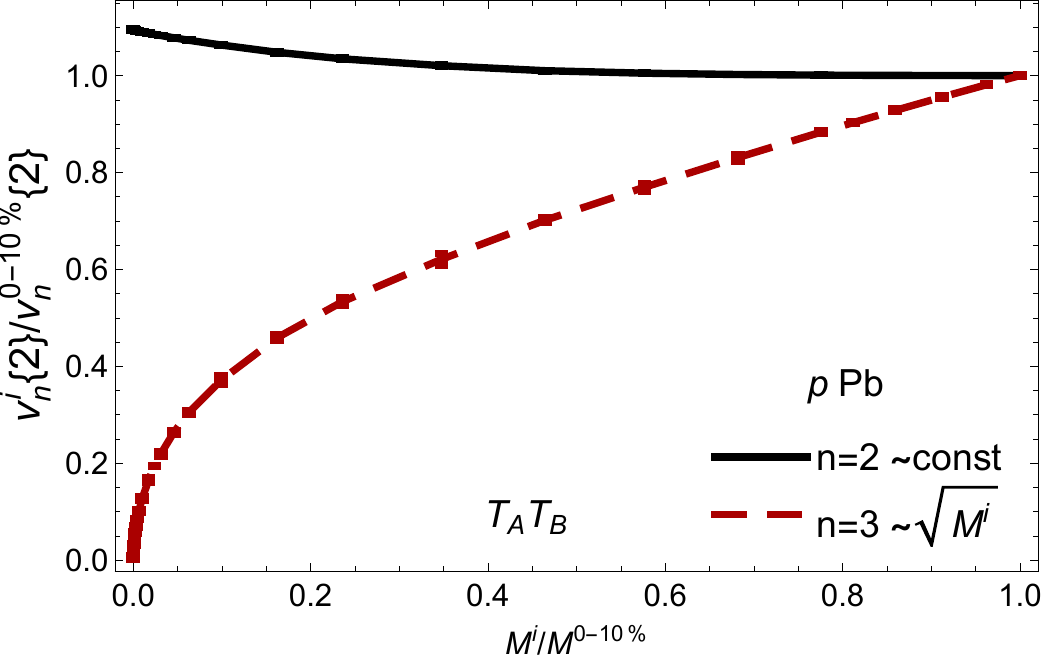}
	\end{subfigure}
	\begin{subfigure}[c]{0.4\linewidth}
		\includegraphics[width=\textwidth]
		{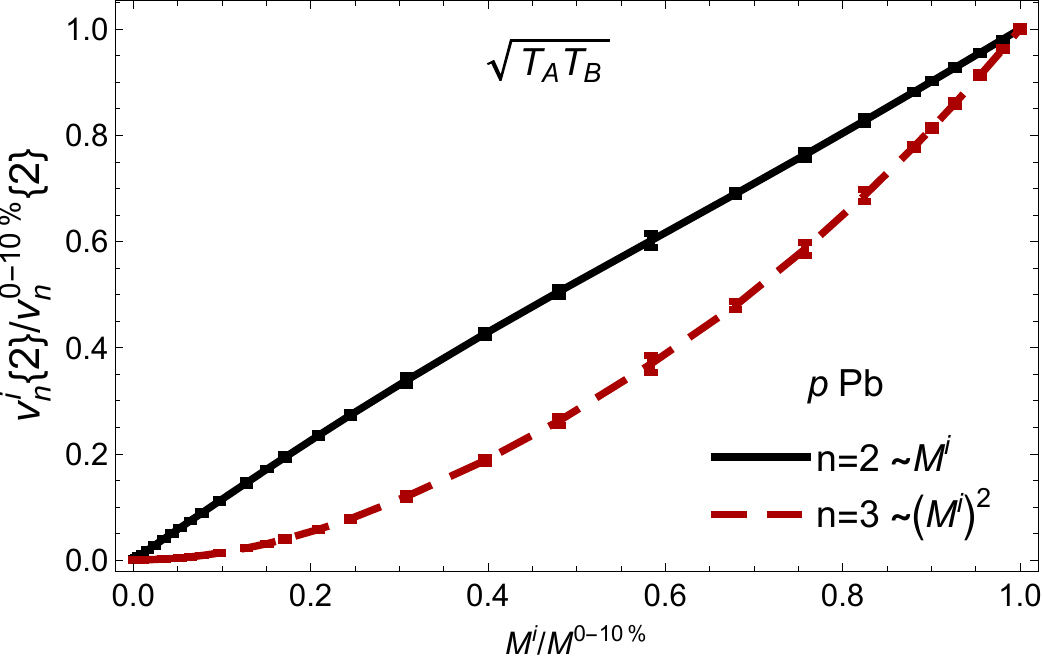}
	\end{subfigure}
	\caption{Elliptic flow ($n=2$, black) and triangular flow ($n=3$, red) versus multiplicity across the entire centrality range for an idealized model of $\XX{}{p}{}{Pb}$ collisions.  As described in the text, a proton of uniform density randomly strikes a smooth optical Glauber profile of the nucleus.  We compare the linear $T_A T_B$ (left panel) versus $\sqrt{T_A T_B}$ (right panel) multiplicity scalings.  Note that the axes are scaled by the most central $0-10\%$ bin in this case due to lower statistics.}
	\label{f:pPbWSTaTb}
\end{figure*}

\begin{figure*}
	\centering
	\begin{subfigure}[c]{0.18\linewidth}
		\includegraphics[width=\textwidth]
		{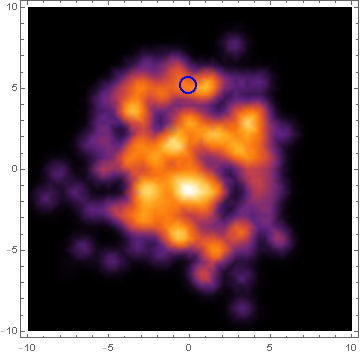}
	\end{subfigure}
	\begin{subfigure}[c]{0.4\linewidth}
		\includegraphics[width=\textwidth]
		{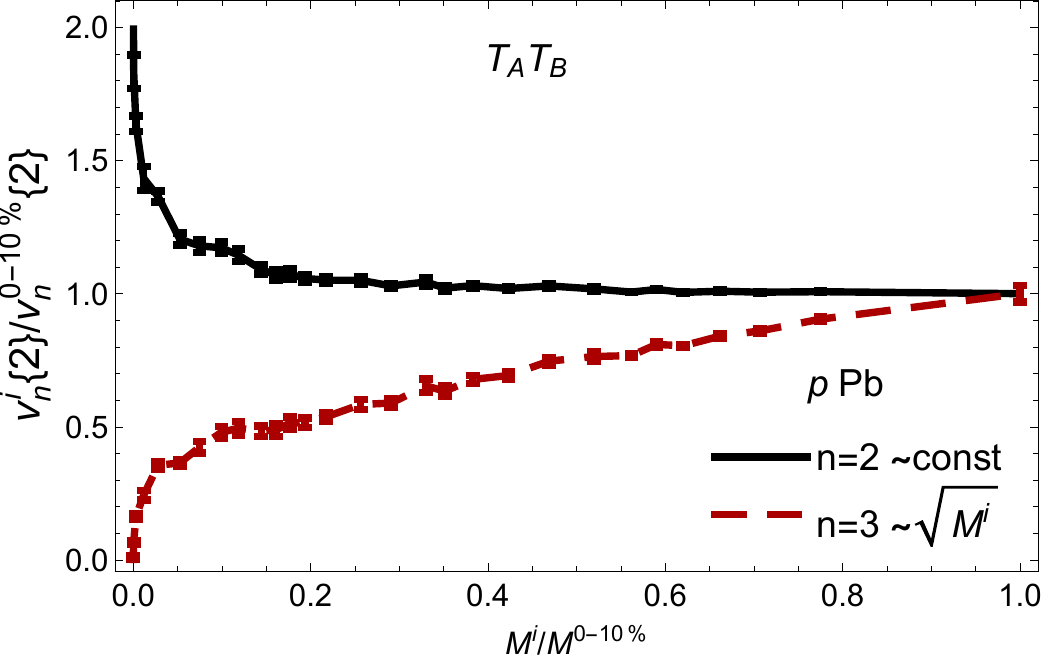}
	\end{subfigure}
	\begin{subfigure}[c]{0.4\linewidth}
		\includegraphics[width=\textwidth]
		{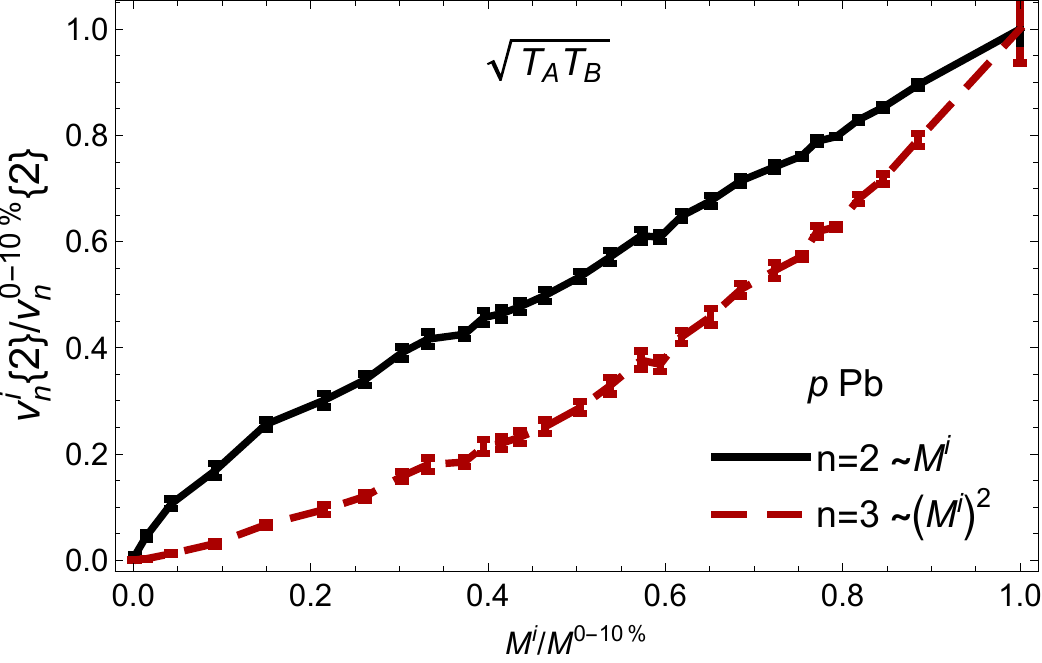}
	\end{subfigure}
	\caption{Same as Fig.~\ref{f:pPbWSTaTb}, but for a constant-density proton on a lumpy nuclear profile generated by Monte Carlo sampling of nucleon positions in the Pb ion.}
	\label{f:pPbMC}
\end{figure*}

To best realize the approximation of a slowly varying geometry inherent in the scalings \eqref{e:CGCmult}, we consider a highly idealized approximation to $\XX{}{p}{}{Pb}$ collisions: a ``cylindrical'' proton of constant density $T_A (\vec{x}_\bot) \propto \theta( 0.5~\mathrm{fm} - | \vec{x}_\bot |)$ which strikes a smooth (optical Glauber) Woods-Saxon profile.  We sample random impact parameters of the proton on the target Pb ion, leading to the results shown in Fig.~\ref{f:pPbWSTaTb}.  In this scenario of artificially slow geometry dependence, we see that the scalings \eqref{e:CGCmult} are realized beautifully.  For linear $T_A T_B$ multiplicity dependence, $v_2 \{2\}$ is roughly constant, and $v_3 \{2\}$ has a downward curvature consistent with a square-root dependence, as anticipated by Ref.~\cite{Mace:2018yvl}.  Likewise, the analogous scaling behavior for the $\sqrt{T_A T_B}$ multiplicity dependence is also clearly satisfied: a linear growth of $v_2 \{2\}$, and an upward curvature for $v_3 \{2\}$ consistent with a quadratic multiplicity dependence.  It is interesting to note, however, that for $T_A T_B$ scaling in peripheral collisions where the density gradients are not negligible, a small deviation from $v_2\{2\} \sim \mathrm{const}$ is seen, giving a slight negative slope to $v_2\{2\}$.

Relaxing these restrictive assumptions about smooth collision geometry in $\XX{}{p}{208}{Pb}$ leads to a smearing of the scalings \eqref{e:CGCmult}, but leaves the overall trends intact.  In Fig.~\ref{f:pPbMC} the constant-density proton is randomly incident on a nuclear profile for the Pb ion which has been generated through Monte Carlo sampling of the nucleon positions.  This effect leads to a slight smearing of the scalings \eqref{e:CGCmult} in most cases, but causes a pronounced deviation of $v_2 \{2\}$ for $T_A T_B$ scaling in peripheral collisions.  For the most part, though, because the collision region set by the constant-density proton, the gradient-based deviations from the scalings \eqref{e:CGCmult} are small.  

\begin{figure*}
	\centering
	\begin{subfigure}[c]{0.18\linewidth}
		\includegraphics[width=\textwidth]
		{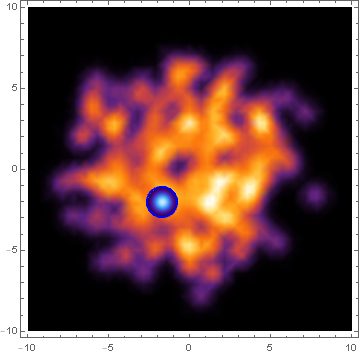}
	\end{subfigure}
	\begin{subfigure}[c]{0.4\linewidth}
		\includegraphics[width=\textwidth]
		{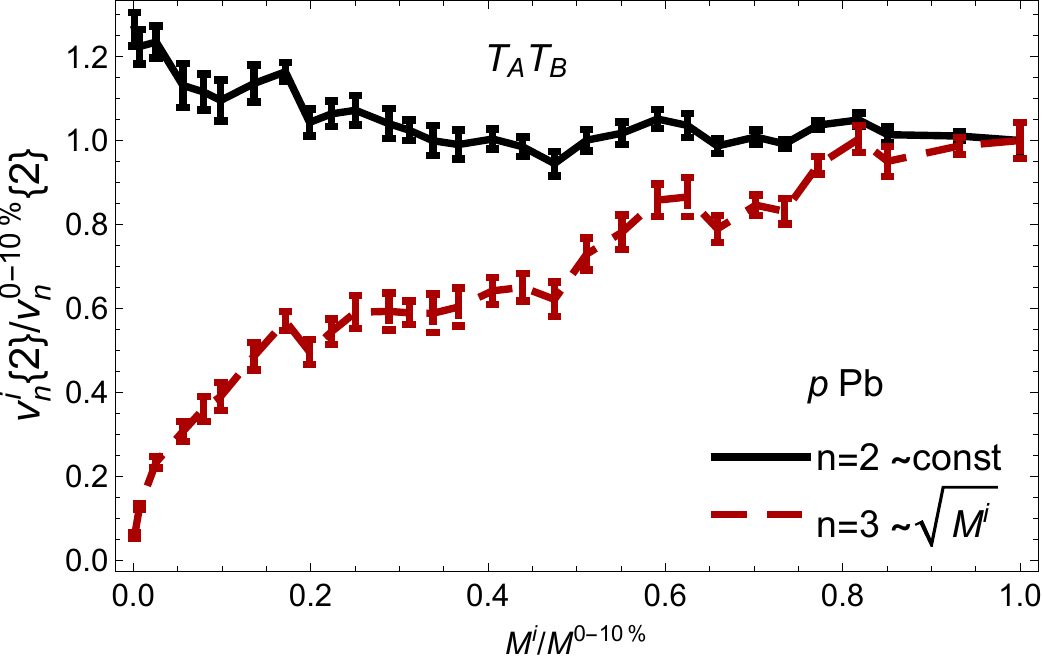}
	\end{subfigure}
	\begin{subfigure}[c]{0.4\linewidth}
		\includegraphics[width=\textwidth]
		{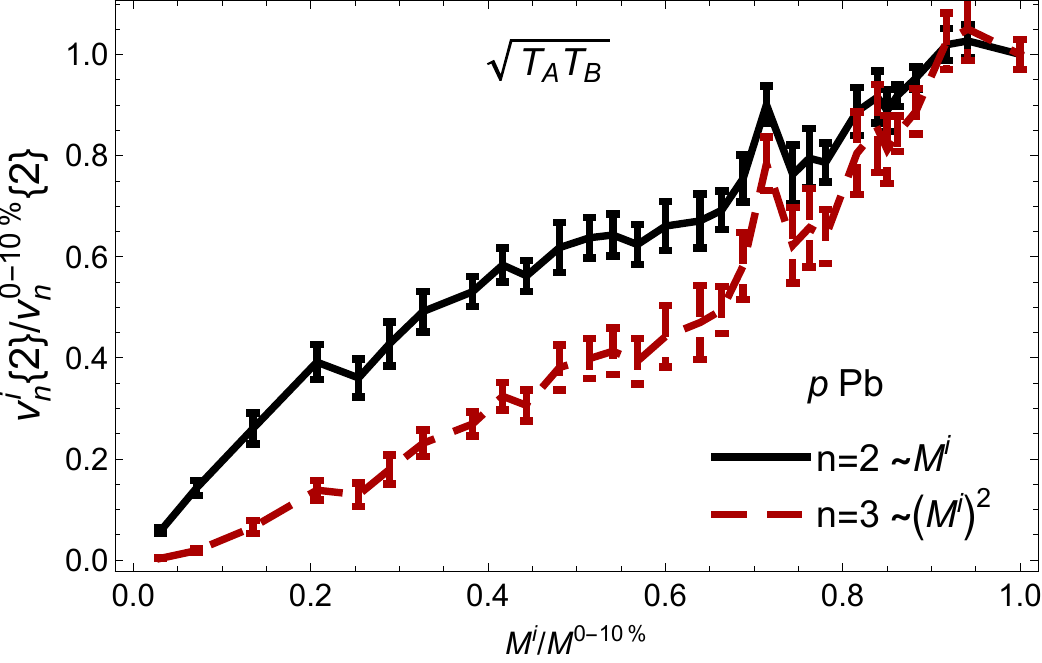}
	\end{subfigure}
	\caption{Same as Fig.~\ref{f:pPbWSTaTb}, but for a proton with a Gaussian profile on a lumpy nuclear profile generated by Monte Carlo sampling of nucleon positions in the Pb ion.}
	\label{f:pPbGaussMC}
\end{figure*}

These deviations become more severe, however, when we also allow the proton to have a nontrivial Gaussian profile of its own, as shown in Fig.~\ref{f:pPbGaussMC}.  Now the structure of the proton samples a wider region because of the tail of its Gaussian distribution, and over this larger region the gradient effects become more significant.  This leads not only to a broadening of the underlying scalings \eqref{e:CGCmult}, but also to a much wider region for which $v_2 \{2\}$ for $T_A T_B$ scaling has a negative slope.

\begin{figure*}
	\centering
	\begin{subfigure}[c]{0.18\linewidth}
		\includegraphics[width=\textwidth]
		{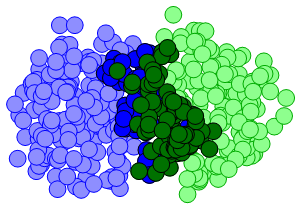}
	\end{subfigure}
	\begin{subfigure}[c]{0.4\linewidth}
		\includegraphics[width=\textwidth]
		{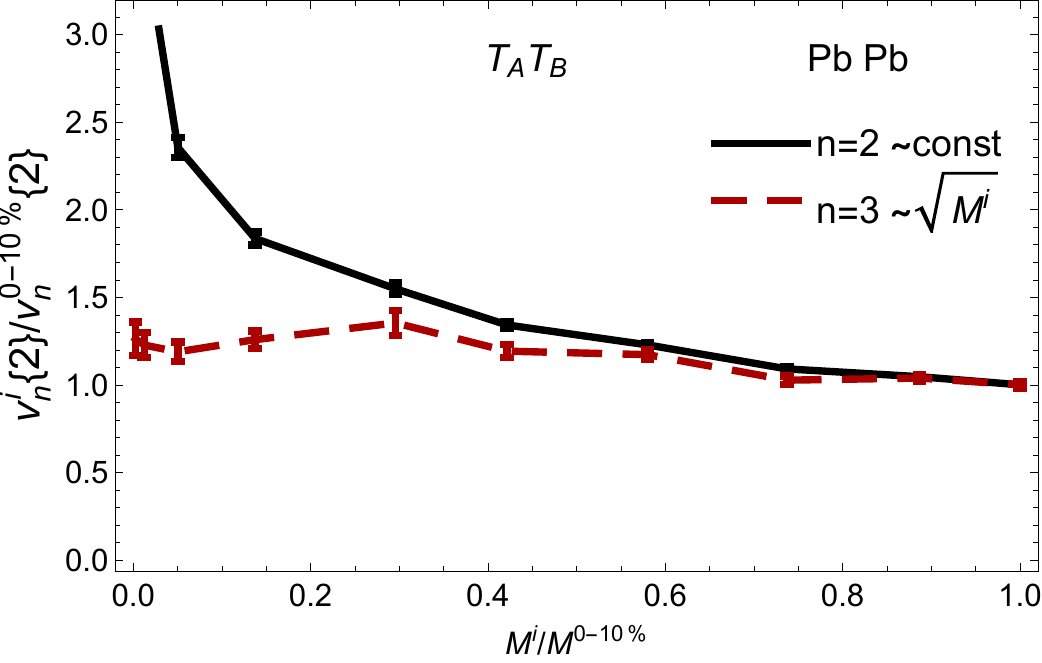}
	\end{subfigure}
	\begin{subfigure}[c]{0.4\linewidth}
		\includegraphics[width=\textwidth]
		{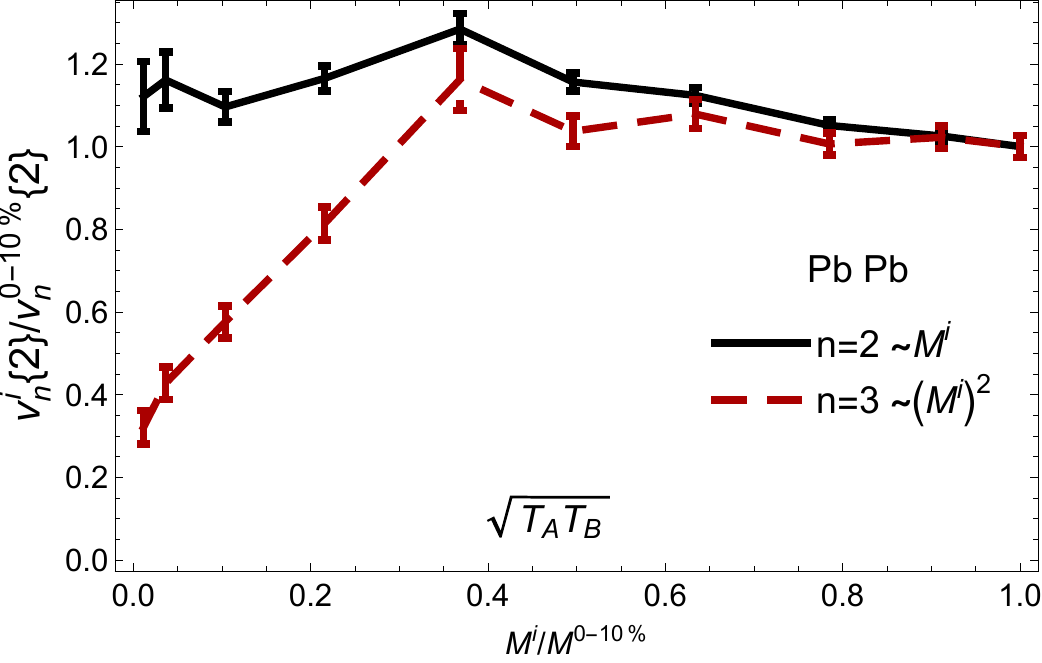}
	\end{subfigure}
	\caption{A simpler model of $\XX{208}{Pb}{208}{Pb}$ collisions than used in TRENTO illustrating the generic flattening of the multiplicity dependence seen in collisions of large, round ions.}
	\label{f:PbPbMc}
\end{figure*}

In Fig.~\ref{f:PbPbMc} we perform a simple simulation of $\XX{208}{Pb}{208}{Pb}$ collisions which uses Monte Carlo sampling of both nuclei and determines the participant geometry by hard-sphere collision detection of the nucleons.  This is still at a level which is far simpler than TRENTO, which uses a more sophisticated scheme for collision detection and includes additional features such as event-by-event multiplicity fluctuations for each participant nucleon.  Here the deviations from the expected scalings \eqref{e:CGCmult} are maximized due to the large scale of the interaction region compared to the gradients of the collision profile, which is itself a complicated combination of the two nuclear densities.  Qualitatively, the large density gradients lead to a significant flattening of the observables, including $v_3 \{2\}$ for $T_A T_B$ scaling and both $v_2 \{2\}$ and $v_3 \{2\}$ for $\sqrt{T_A T_B}$ scaling.  It is notable, however, that $v_2 \{2\}$ for $T_A T_B$ scaling now exhibits a negative slope over almost the entire multiplicity range due to the gradients.  We thus conclude that a large collision region over a very lumpy collision profile generally results in a flattening of the flow harmonics in a CGC picture, which in the extreme can completely damp out the baseline scalings in Eqs.~\eqref{e:CGCmult}.

The multiplicity scalings seen here for smooth realizations of $\XX{}{p}{208}{Pb}$ collisions, and the deviations from them with more realistic collision geometries and for simple $\XX{208}{Pb}{208}{Pb}$ collisions, show the baseline behavior one can expect from the CGC correlations in the absence of nuclear deformations.  In contrast, for the same observables a different scaling was shown in Ref.~\cite{Kovchegov:2013ewa} for collisions of deformed ions such as $\XX{238}{U}{238}{U}$.  For a smooth optical Glauber profile with ellipsoidal geometry
\begin{align}
\rho(\vec{r}) = \rho_0 \exp\left[ - \frac{x^2 + y^2}{R^2} - \lambda^2 \frac{z^2}{R^2} \right]
\end{align}
the ratio of tip-on-tip to side-on-side collision geometries yields
\begin{align} \label{e:Dougdeform}
\frac{ \left[ v_2 \{2\} \right]_{\mathrm{tip}} }
{\left[ v_2 \{2\} \right]_{\mathrm{side}} } =
\frac{1}{\lambda} \approx 1.26
\end{align}
where $T_A T_B$ scaling has been assumed and $\lambda \approx 0.79$ is reasonable for parameterizations of ${}^{238} \mathrm{U}$.  

\begin{figure*}
	\centering
	\begin{subfigure}[c]{0.18\linewidth}
		\includegraphics[width=\textwidth]
		{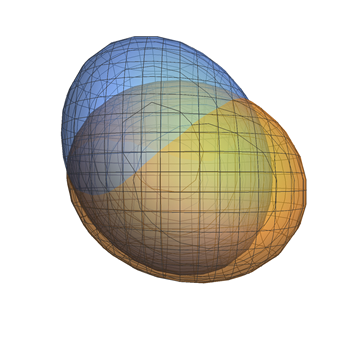}
	\end{subfigure}
	\begin{subfigure}[c]{0.4\linewidth}
		\includegraphics[width=\textwidth]
		{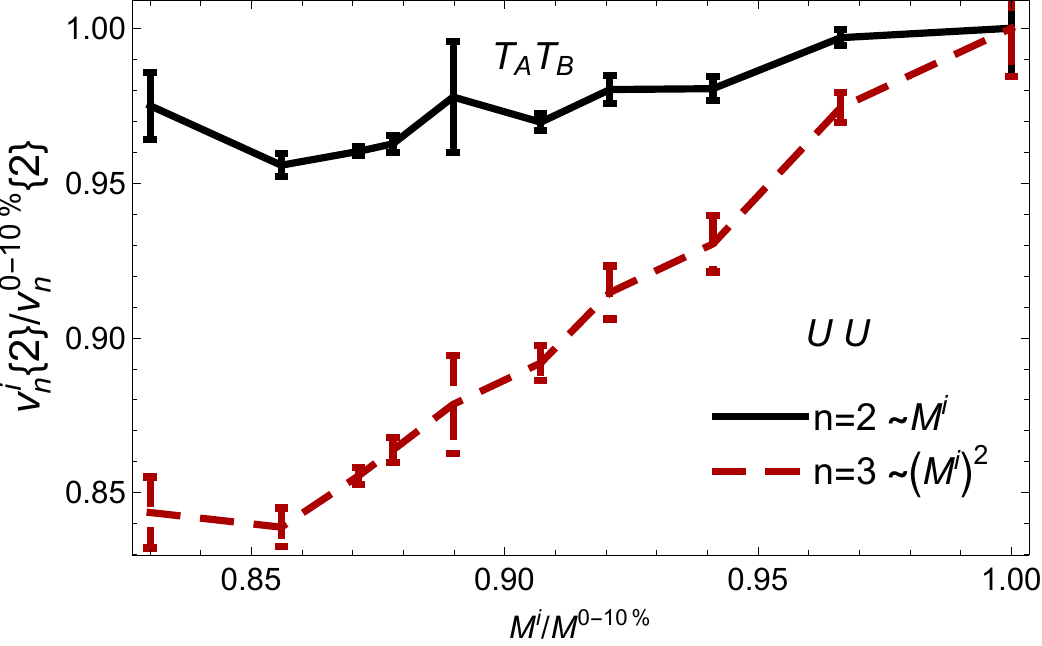}
	\end{subfigure}
	\begin{subfigure}[c]{0.4\linewidth}
		\includegraphics[width=\textwidth]
		{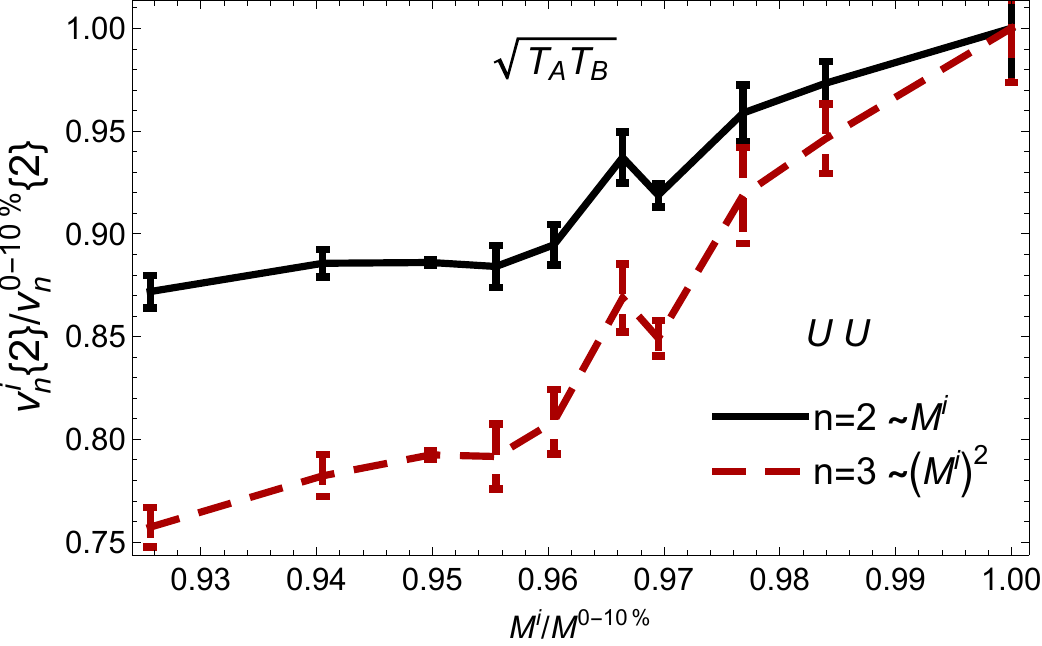}
	\end{subfigure}
	\caption{Smooth optical Glauber $\XX{238}{U}{238}{U}$ collisions realizing the different scaling \eqref{e:Dougdeform} for deformed nuclei.}
	\label{f:UUsmooth}
\end{figure*}

Thus the multiplicity scalings \eqref{e:CGCmult} arising from a smooth round geometry seen in Fig.~\ref{f:pPbWSTaTb} are modified by nuclear deformation, leading instead to $v_2\{2\}$ which increases with multiplicity instead of remaining constant.  This prediction is realized in Fig.~\ref{f:UUsmooth} for smooth optical Glauber collisions of $\XX{238}{U}{238}{U}$.  Here we see that, because of the deformation, all of the curves now show an increasing and roughly linear trend with the multiplicity.  Most importantly, $v_2\{2\}$ for $T_A T_B$ scaling has deviated from its constancy to show a positive slope for the first time due to the deformation.

\begin{figure*}
	\centering
	\begin{subfigure}[c]{0.18\linewidth}
		\includegraphics[width=\textwidth]
		{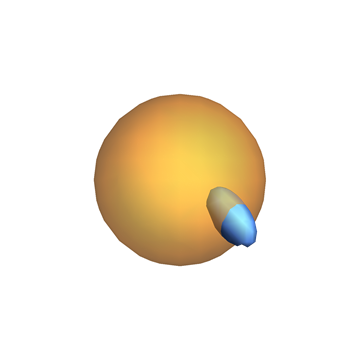}
	\end{subfigure}
	\begin{subfigure}[c]{0.4\linewidth}
		\includegraphics[width=\textwidth]
		{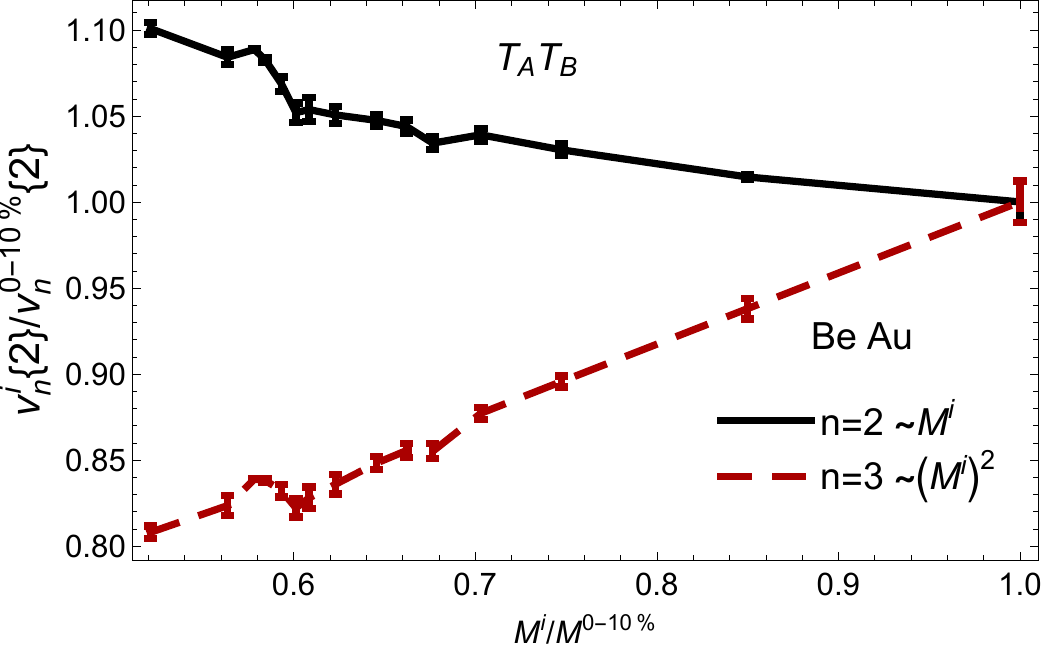}
	\end{subfigure}
	\begin{subfigure}[c]{0.4\linewidth}
		\includegraphics[width=\textwidth]
		{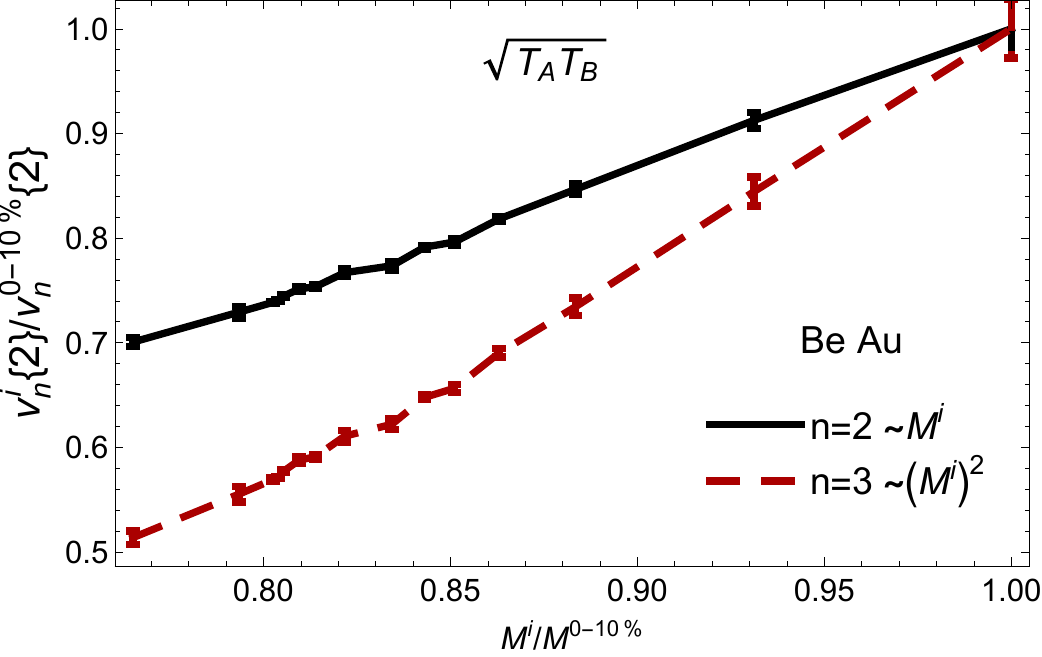}
	\end{subfigure}
	\caption{Smooth optical $\XX{9}{Be}{197}{Au}$ collisions exploring the role of nuclear deformations in smaller, asymmetric collisions.}
	\label{f:BeAusmoothdef}
\end{figure*}

Finally, in Fig.~\ref{f:BeAusmoothdef} we similarly show the modification of the scalings \eqref{e:CGCmult} due to nuclear deformation for the asymmetric collision of a small deformed ion on a larger round one.  Here we use smooth optical Glauber profiles for $\XX{9}{Be}{197}{Au}$ collisions as a way to explore an intermediate collision system.  Most of the trends seen for smooth $\XX{238}{U}{238}{U}$ collisions are already present in Fig.~\ref{f:BeAusmoothdef} for $\XX{9}{Be}{197}{Au}$: flow harmonics which increase linearly with multiplicity in most cases.  The exception is that we still see a small negative slope of $v_2 \{2\}$ with $T_A T_B$ scaling, similar to what was seen for the collision of undeformed systems previously.   We thus suspect that this is because we are not yet seeing the full impact of the deformation in a truly ultracentral regime.  

This analysis demonstrates that the multiplicity scaling $v_2\{2\} \propto N_{\mathrm{ch}}^0$ and $v_3\{2\} \propto N_{\mathrm{ch}}^{1/2}$ for multiplicities computed proportional to $T_A T_B$ in the CGC are well-realized in $\XX{}{p}{208}{Pb}$ collisions, even when a lumpy nuclear profile on the scale of the proton is present.  These scalings are violated, however, when the collision region encompasses regions having significant gradients, such as in heavy-ion collisions like $\XX{208}{Pb}{208}{Pb}$ and presumably as well for smaller systems with sub-nucleonic structure.  And importantly, even for smooth density profiles, these scalings can change significantly due to the presence of deformed nuclei, leading in particular to a $v_2\{2\}$ which grows with multiplicity in ultracentral collisions, as first predicted in Ref.~\cite{Kovchegov:2013ewa}.

\section*{References}
%\bibliography{BIG}

%

\end{document}